\documentclass[sigconf]{acmart}
\renewcommand\footnotetextcopyrightpermission[1]{} 
\copyrightyear{2020}
\acmYear{2020}
\setcopyright{acmcopyright}\acmConference[MODELS '20]{ACM/IEEE 23rd International Conference on Model Driven Engineering Languages and Systems}{October 18--23, 2020}{Montreal, QC, Canada}
\acmBooktitle{ACM/IEEE 23rd International Conference on Model Driven Engineering Languages and Systems (MODELS '20), October 18--23, 2020, Montreal, QC, Canada} \acmPrice{15.00}
\acmDOI{10.1145/3365438.3410961}
\acmISBN{978-1-4503-7019-6/20/10}

\AtBeginDocument{%
	\providecommand\BibTeX{{%
			\normalfont B\kern-0.5em{\scshape i\kern-0.25em b}\kern-0.8em\TeX}}
}

\usepackage{preamble}
\begin{document}
\title[A Scalable Querying Scheme for Memory-efficient Runtime Models with History]{A Scalable Querying Scheme for \\ Memory-efficient Runtime Models with History}
\author{Lucas Sakizloglou,
Sona Ghahremani,
Matthias Barkowsky, and
Holger Giese}
\email{<first-name>.<last-name>@hpi.de}
\affiliation{%
  \institution{Hasso Plattner Institute, University of Potsdam, Germany}
}
\renewcommand{\shortauthors}{Sakizloglou et al.}
\begin{abstract}
Runtime models provide a snapshot of a  system at runtime at a desired level of abstraction. Via a causal connection to the modeled system and by employing model-driven engineering techniques, runtime models support schemes for (runtime) adaptation where data from previous snapshots facilitate more informed decisions. Nevertheless, although runtime models and model-based adaptation techniques 
have been the focus of extensive research, schemes that treat the evolution of the model over time as a first-class citizen have only lately received attention. Consequently, there is a lack of sophisticated technology for such runtime models with history.

We present a querying scheme where the integration of temporal requirements with incremental model queries enables scalable querying for runtime models with history. Moreover, our scheme provides for a memory-efficient storage of such models. By integrating these two features into an adaptation loop, we enable efficient history-aware self-adaptation via runtime models, of which we present an implementation.
\end{abstract}
\maketitle
%
\section{Introduction}\label{sec:intro}
A runtime model provides a view on a running system at a desired level of abstraction that can be used for monitoring, analyzing, or adapting the system through a causal connection between the model and the system~\cite{DBLP:conf/models/VogelSG10}, i.e., any relevant change of the system is reflected in the model and vice versa~\cite{Blair+2009}. 
Runtime models typically capture snapshot-based representations of the modeled system in its current state~\cite{bencomo2014models}. Thereby, they provide an abstract view on a current system configuration that, via causal connection and  employing model-driven engineering techniques, can support online (model-based) 
adaptation schemes~\cite{Blair+2009, Ghahremani.2017.Efficient} which mitigate the difficulty of managing complex interconnected systems~\cite{MANOGARAN2018375}.

Capturing the evolution of runtime models
~\cite{artale2007evolving} has been shown to be a promising direction to cope with the increasing complexity of software systems and their dynamic environments~\cite{Bencomo.2019.Models}. Furthermore, it is often desired, and sometimes required by application domains, e.g.,  healthcare ~\cite{Combi_2012_ModellingTemporalDatacentricMedicalProcesses}, that model-based schemes recollect previous observations and utilize these \emph{historical data} in future activities: For instance, more informed adaptations can be enabled via expanding  runtime models to capture 
the history of system changes and  interactions~\cite{esfahani2016inferring}, which  can be utilized to address emergent circumstances ~\cite{sebestyenova2007case} or predict potential future changes~\cite{moreno_2015_ProactiveSelfAdaptation}.

Although runtime models, model queries, and adaptation based on runtime model changes have been the focus of extensive research (cf.~\cite{DBLP:journals/sosym/BencomoGS19}), schemes that treat the evolution of the runtime model over time, referred to as \emph{Runtime Model with History} (\RTMwH) in the following, as a first-class citizen have only lately received attention (cf.~\cite{Bencomo2019}). Moreover, in order to utilize the history of complex 
systems that operate in highly dynamic environments for online adaptations, \RTMwH technology should be capable of consolidating numerous changes into the \RTMwH, often arriving at a high pace ~\cite{DBLP:journals/iotj/CatarinucciDMPP15} and in the form of events~\cite{David_2018_FoundationsforStreamingModelTransformationsbyComplexEventProcessing}, as well as provide facilities for storing and querying the historical event data in a scalable manner.

Runtime models have been  utilized in (self-)adaptation schemes where incremental model queries are employed to detect issues requiring system adaptation, e.g., failures, in an efficient manner (cf.~\cite{SG-TAAS}). However, the history of system changes is not captured and adaptation decisions are only made based on the current system state. 
The idea of runtime models enriched with history in the form of past event data where queries impose temporal requirements on matched patterns has been, so far, treated only preliminarily and ad hoc, e.g., by a manual translation of a single example of a restricted form that supports only past requirements~\cite{2020_towards_highly_scalable_runtime_models_with_history}. In this paper, we extend the 
scheme envisioned in ~\cite{2020_towards_highly_scalable_runtime_models_with_history} and present a version 
which lifts the restrictions, supports complex queries with both past and future requirements, and their systematic operationalization. 
We extend the scheme further to support a full online adaptation cycle employing incremental model queries, which enables history-aware adaptations where real-time, efficient storage and querying for data generated by events are key.

Our contributions are as follows. First, we present a scalable online scheme for the incremental processing of pattern-based model queries which support temporal logic operators. The approach automatically maps a temporal graph logic formula to a network of simple graph sub-queries. 
Secondly, our scheme allows for memory-efficient \RTMwH via an automated, a priori analysis of the model queries that only keeps data in the \RTMwH that are necessary to evaluate the model queries correctly.
By integrating these two contributions into a self-adaptation loop, we enable efficient history-aware self-adaptation via runtime models. 
Finally, we present an implementation of the querying scheme embedded in an adaptation loop, evaluate it on simulated real and synthetic logs from the medical domain, and compare it to a baseline acquired by a relevant state-of-the-art tool.

The rest of the paper is organized as follows. ~\autoref{sec:foundations} discusses the building blocks of our scheme. An overview of the scheme
and its utilization to enable history-aware self-adaptation is presented in~\autoref{sec:overview}. 
The incremental matching of patterns with temporal requirements is presented in~\autoref{sec:approach}, while~\autoref{sec:approach2} details the query analysis that enables memory-efficient \RTMwH. We evaluate the performance of our prototypical implementation in~\autoref{sec:evaluation}, discuss related work in~\autoref{sec:related}, and conclude the paper as well as discuss future work in~\autoref{sec:conclusion}.
\section{Foundations}
\label{sec:foundations}

\begin{figure}
	\centering
		\includegraphics[width=\linewidth]{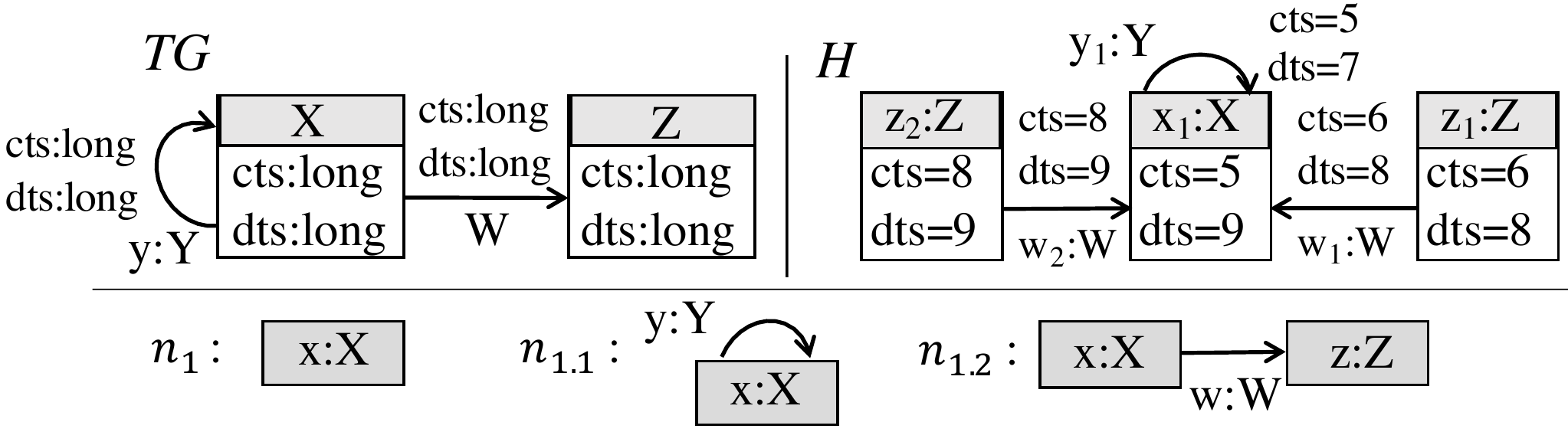}
	\caption{\label{fig:basic}
		Type graph \emph{TG} of $\Sigma$, the \RTMwH $H$, and $n$ patterns}
\end{figure}

\subsection{Runtime Models for Self-adaptation}\label{subsec:RTM}
\label{subsec:graph-based-models}
A \emph{runtime model} captures a snapshot-based representation of the current state of the modeled system at a desired level of abstraction~\cite{bencomo2014models}. Self-adaptation can be generally achieved by adding, removing, and re-configuring components as well as connectors among components in the system architecture~\cite{MageeKramer1996}, therefore, software architecture is typically considered an appropriate abstraction level, e.g.,~\cite{Garlan+2009,Ghahremani.2017.Efficient}.
Runtime models may be used for adapting the system through a causal connection between the model and the system~\cite{DBLP:conf/models/VogelSG10}.
\emph{Model queries} are employed to retrieve data from a runtime model. The established practice of representing the runtime model as a graph captures architectural components as vertices, connectors between components as edges, and information about the components as attributes~\cite{VG10}. The model conforms to a \emph{metamodel} that specifies a language for runtime models and defines types of vertices, edges, and attributes. 

Formally, a graph-based runtime model can be represented as a \emph{typed attributed graph} where a graph is typed over a \emph{type graph}, in the same manner a runtime model conforms to its metamodel. For an example, see \autoref{fig:basic} for the type graph $TG$  of an abstract elementary system $\Sigma$. 
A graph-based representation allows for the utilization of established formalisms, such as \emph{typed, attributed graph transformation}~\cite{DBLP:conf/gg/EhrigPT04}, for the maintenance and adaptation of the model, whereby \emph{graph transformation rules} are employed to capture model queries as well as perform in-place model transformations. 

In short, let $G$ be a graph representation of the runtime model (effectively, the system architecture), 
and $\rho$ a graph transformation rule. A rule $\rho$ is characterized by a left-hand side (\emph{LHS}) and a right-hand side (\emph{RHS}) graph pattern which define the precondition and postcondition of an application of $\rho$ respectively. In this context, the \emph{LHS} describes a structural fragment of the architecture and the \RHS the corresponding model transformation. A match $m$ of \LHS in $G$ corresponds to an occurrence of \LHS in $G$ and identifies a part of the runtime model where the transformation should occur. The \LHS of a rule can also be used to characterize a \emph{graph query}, which is the equivalent graph-based notion of a model query. 

To realize architectural self-adaptation, a system is equipped with a \textit{MAPE-K} feedback loop that \underline{m}onitors and \underline{a}nalyzes the system and, if needed, \underline{p}lans and \underline{e}xecutes an adaptation of the system via making architectural changes, i.e., adding and removing components as well as connectors among components in the system architecture~\cite{Garlan+2009}. All four MAPE activities are based on \underline{k}nowledge~\cite{Kephart&Chess2003}. 
The feedback loop maintains a runtime model as part of its knowledge to represent the \textit{current state} of the architecture. Rule-based self-adaptation schemes employ \emph{adaptation rules} to capture events~(during monitoring phase), check whether the events triggered any adaptation issues~(during analysis) and plan for and execute an adaptation~(during planning and execution respectively)~\cite{Rule-based_SASLanese2010}. The graph-based representation of runtime models allows for a realization of adaptation rules in form of graph transformation rules where analysis is performed via model queries and the system is adapted via in-place model transformations~\cite{Ghahremani.2017.Efficient}. 

\subsection{Efficient Pattern Matching for Queries}
\label{subsec:query-operationalization}
The process of finding matches of \LHS patterns in $G$ is called \emph{graph pattern matching} and corresponds to the execution of a graph query specified by the pattern \LHS. In certain cases however, simple patterns are not sufficient as a language 
for specifying more sophisticated \emph{application conditions}
of adaptation rules, for instance if the existence of certain model elements should be prohibited. In those cases, \LHS patterns and thus graph queries are enhanced with a set of application conditions \ac which every match $m$ should satisfy. In the following, a graph query $q$ is characterized by a pattern $n$ and application conditions \ac, denoted $q(n,\,$\ac). 

\begin{figure}[t]
	\centering
		\includegraphics[width=\linewidth]{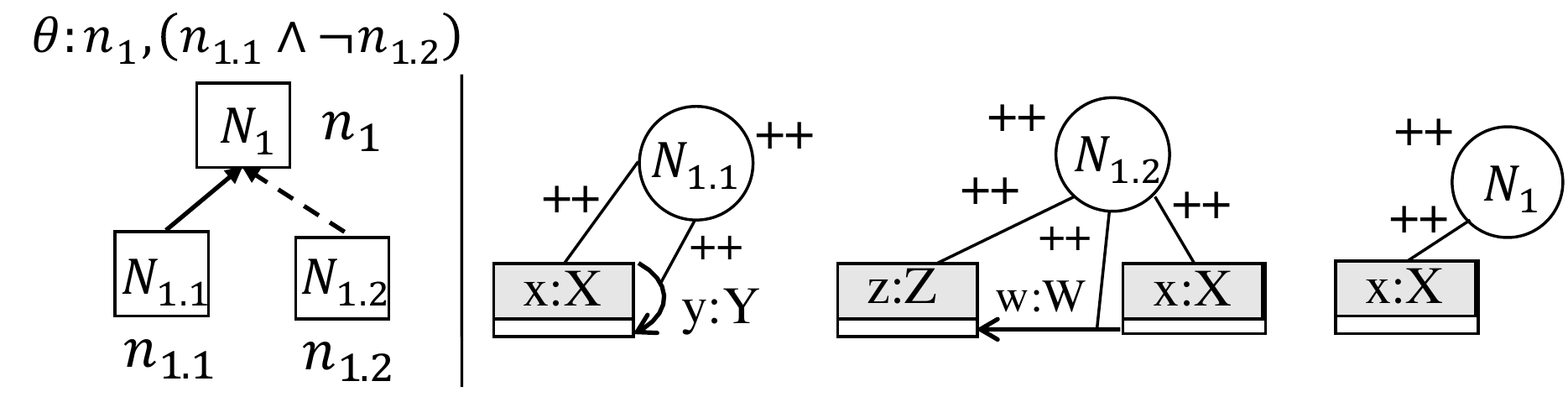}
	\caption{\label{fig:gdn}
		Example: GDN (left) and Marking Rules for $\theta$}
\end{figure}

A graph query is a declarative means to express a sought pattern and its application conditions. The query itself does not specify a method for its \emph{operationalization}, i.e., instructions on how to execute the query over a graph. In this work, for the operationalization of queries, we build on ~\cite{DBLP:conf/gg/BeyhlBGL16}, which supports \ac formulated as \emph{Nested Graph Conditions} (\NGCs)~\cite{Habel_2009_Correctnessofhighleveltransformationsystemsrelativetonestedconditions}. \NGCs support the nesting of patterns to bind graph elements in outer conditions and relate them to inner (nested) conditions. Moreover, \NGCs support the first-order logic operators negation ($\neg$), existential quantification ($\exists$), and conjunction ($\wedge$) and thus have the expressive power of first-order logic on graphs~\cite{Rensink_2004_Representingfirstorderlogicusinggraphs} and constitute, as such, a natural formal foundation for pattern-based queries. The approach in~\cite{DBLP:conf/gg/BeyhlBGL16} presents a formal framework for the decomposition of a query with an arbitrarily complex \NGC as \ac into a suitable ordering of simpler sub-queries, which is called a \emph{generalized discrimination network} (\GDN). A \GDN is a directed acyclic graph where each graph node represents a (sub-)query. To avoid confusion, we refer to the GDN as a \emph{network}. Dependencies between sub-queries are represented by edges from child nodes, i.e. the nodes whose results are required, to the parent node, i.e. the node which requires the results. Dependencies can either be \emph{positive}, i.e.  the sub-query realized by the parent node requires the presence of matches of the child node, or \emph{negative}, i.e., the sub-query of the parent node forbids the presence of such matches.
The query is executed bottom-up: the execution starts with leaves and proceeds upwards in the network. 
The terminal node computes the overall result of the query.

In~\cite{DBLP:conf/gg/BeyhlBGL16}, a GDN is realized as a set of  graph transformation rules where each GDN node, i.e., each (sub-)query, is associated with one transformation rule. The \LHS of the rule searches for matches of the corresponding query in a given graph $G$. The \RHS of the rule creates for each match a \emph{marking node} in $G$ and \emph{marking edges} from the marking node to each element of the match---marking nodes are not to be confused with regular graph nodes in $G$ (which, in this context, represent system components) thus we use the term \emph{vertex} for the latter.
In order to be able to create marking nodes and edges, the transformation rules of a \GDN are typed over an extended type graph which adds the required types for marking nodes and edges to the initial type graph.
The \LHS of queries with dependencies have \ac that require the existence of marking nodes of their positive dependencies and forbid the existence of marking nodes of their negative dependencies.

As an example, assume the following graph query 
$q(n_1,\theta)$ where $\theta \coloneqq n_1, (n_{1.1}  \wedge \neg n_{1.2})$, which captures the following (structural) requirement: all matches of $ n_1$ can be extended to a pattern $n_{1.1}$ but not to a pattern $n_{1.2}$ The patterns are based on the $TG$ of system $\Sigma$ (introduced in~\autoref{subsec:RTM}) and are shown  in~\autoref{fig:basic} (bottom). The nesting of patterns implies a \emph{binding} of the vertex of type X from inner patterns, i.e., all patterns refer to the same element in $H$. Note that, for presentation purposes, we adopt a simplified grammar for \NGCs that omits existential quantifiers for patterns as well as the operator $\TOP$ which is always satisfied and marks the end of a nested condition, e.g., in $\theta$, we write $n_{1.1}$ instead of  $\exists(n_{1.1}, \TOP)$.

The \GDN for $q$ is shown in~\autoref{fig:gdn} (left), where each square represents a GDN node. Each node is associated with a marking rule.
The \GDN consists of three nodes, i.e., rules: the node $N_{1.1}$ for the query searching for $n_{1.1}$, the node $N_{1.2}$ for pattern $n_{1.2}$ and the topmost node $N_{1}$ for pattern $n_1$. Node $N_{1}$ computes its matches by matching its pattern and checking whether both of its dependencies are satisfied (conjunction in $\theta$). The negative dependency which captures the negation in $\theta$ (drawn by a dashed line) is satisfied when a match for $N_{1}$ can not be extended by a match for $N_{1.2}$. All nodes are realized by transformation rules whose \LHS matches a pattern and whose \RHS creates marking nodes and edges that mark the matches of the \LHS. The rules for nodes $N_{1.1}$, $N_{1.2}$, $N_{1}$ are shown in \autoref{fig:gdn}, where (i) vertices are shown by rectangles, (ii) marking nodes by circles, and (iii) the marking nodes and edges added by a rule are annotated with "$++$".
For presentation purposes, the illustrations of rules contain both \LHS and \RHS.

A \GDN is capable of being executed \emph{incrementally} and thus enables efficient, incremental pattern matching. Changes in $G$ can propagate through the network, whose nodes only recompute their results if the change concerns them or one of their dependencies.

\subsection{Runtime Models with History}

A Runtime Model with History (\RTMwH)~\cite{2020_towards_highly_scalable_runtime_models_with_history} consolidates the evolution of a runtime model in a single graph and as such it constitutes the cornerstone of our scheme. An \RTMwH can be obtained by an adjustment to the system metamodel such that each element is equipped with a \emph{creation} and \emph{deletion} timestamp, \emph{cts} and \emph{dts} respectively---see \emph{TG} in \autoref{fig:basic}. Based on monitoring data, when an element is created or deleted in the represented system, its \emph{cts}, respectively \emph{dts}, is updated accordingly in the \RTMwH. At the time of creation, the \emph{dts} of an element is set by default to $\infty$. For an example of a \RTMwH, see \emph{H} in~\autoref{fig:basic}, where the \emph{cts} and \emph{dts} of each element reflects the latest monitoring data. Note that, some elements, e.g., the edge \emph{y}$_1$, have been removed from the modeled system yet featured in the \RTMwH. By featuring removed elements, i.e., components whose \emph{dts} is in the past or present, an \RTMwH  transcends the traditional notion of causal connection. 
\subsection{Queries over \RTMwH}
\label{subsec:queries-over-rtmwh}
We introduce graph queries that, via their \ac, express temporal requirements on patterns. Such queries provide a powerful means to query the history or evolution of a system execution, as the latter is reflected in \RTMwH. An example is the following requirement: 
``\emph{For all matches of $n_1$ in H at a time point $t$, 
at least one match of $n_{1.2}$ should be found at some time point $t' \in [t, t + 2]$, that is at most 2 time units later. In addition, at each time point $t'' \in [t, t')$ in between, at least one match for $n_{1.1}$ should be present.}'', where all patterns refer to the same vertex of type X in \emph{H}.
The specification of such requirements is enabled by \emph{Metric Temporal Graph Logic} (\MTGL)~\cite{Giese_2019_MetricTemporalGraphLogicoverTypedAttributedGraphs, mtgl-journal}. \MTGL builds on \NGCs and \emph{Metric Temporal Logic}~\cite{Koymans_1990_Specifyingrealtimepropertieswithmetrictemporallogic} to enable the specification of \emph{Metric Temporal Graph Conditions} (\MTGCs) which support \emph{metric}, i.e, interval-based, temporal operators: the future \emph{until} ($\mathrm{U}_I$, where $I$ is a time interval\footnote{A (time) interval $I$ is a set  $I = \{t\,|\,t \in \rm I\!R_{0}^{+} \wedge \tau \leq t \leq \tau'\}$, where
$\tau, \tau' \in \rm I\!R_{0}^{+}$, and $\tau, \tau'$ is the \emph{lower}, respectively \emph{upper}, bound of the interval. An interval $I$ is also denoted by $[\tau, \tau']$ and its lower and upper bound by  $\ell(I)$, respectively $u(I)$.}) and \emph{eventually} ($\lozenge_I$), and their dual past operators \emph{since} ($S_I$) and \emph{once} $(\blacklozenge_I)$.

\MTGL reasons over a sequence of graphs, which in this context represents consecutive snapshots of the runtime model. However, as shown in~\cite{Giese_2019_MetricTemporalGraphLogicoverTypedAttributedGraphs}, a graph sequence can be uniquely folded into a \emph{graph with history}. In a graph with history, each node and edge is associated with a creation timestamp \emph{cts} and a deletion timestamp \emph{dts}. To store these values, the type graph is extended by appropriate attributes. \MTGCs can also be equivalently checked over a graph with history, which here corresponds to an \RTMwH.

The exemplary requirement above is captured by the \MTGC $\zeta\coloneqq n_1, ( n_{1.1} \,\mathrm{U}_{[0,2]}\,n_{1.2})$.
The intuition behind \emph{until} is reversed for the past operator \emph{since}, 
e.g.  $(n_{1.1} \, \mathrm{S}_{[0,2]} \, n_{1.2}$),
which requires that when $n_{1.1}$ is matched at time point $t$, a match for $n_{1.2}$ has existed at some time point $t' \in [t-2, t]$, and that at every time point $t'' \in (t', t]$ in between, a match for $n_{1.1}$ is present in the graph. The operators \emph{eventually}  ($\lozenge_I$)  and \emph{once} $(\blacklozenge_I)$ are abbreviations of \emph{until} and \emph{since}: $\lozenge_I\,n_1= \TOP\, \mathrm{U}_I \,n_1$ and $\blacklozenge_I\,n_1=\TOP\,\mathrm{S}_I\,n_1$. The query $q(n_1,\zeta)$ computes all matches of $n_1$ in $H$ that satisfy the \MTGC $\zeta$.
\section{\APPROACHNAME: A Querying Scheme Extended for Self-adaptation}
\label{sec:overview}

\begin{figure}[t]
	\includegraphics[width=0.9\linewidth]{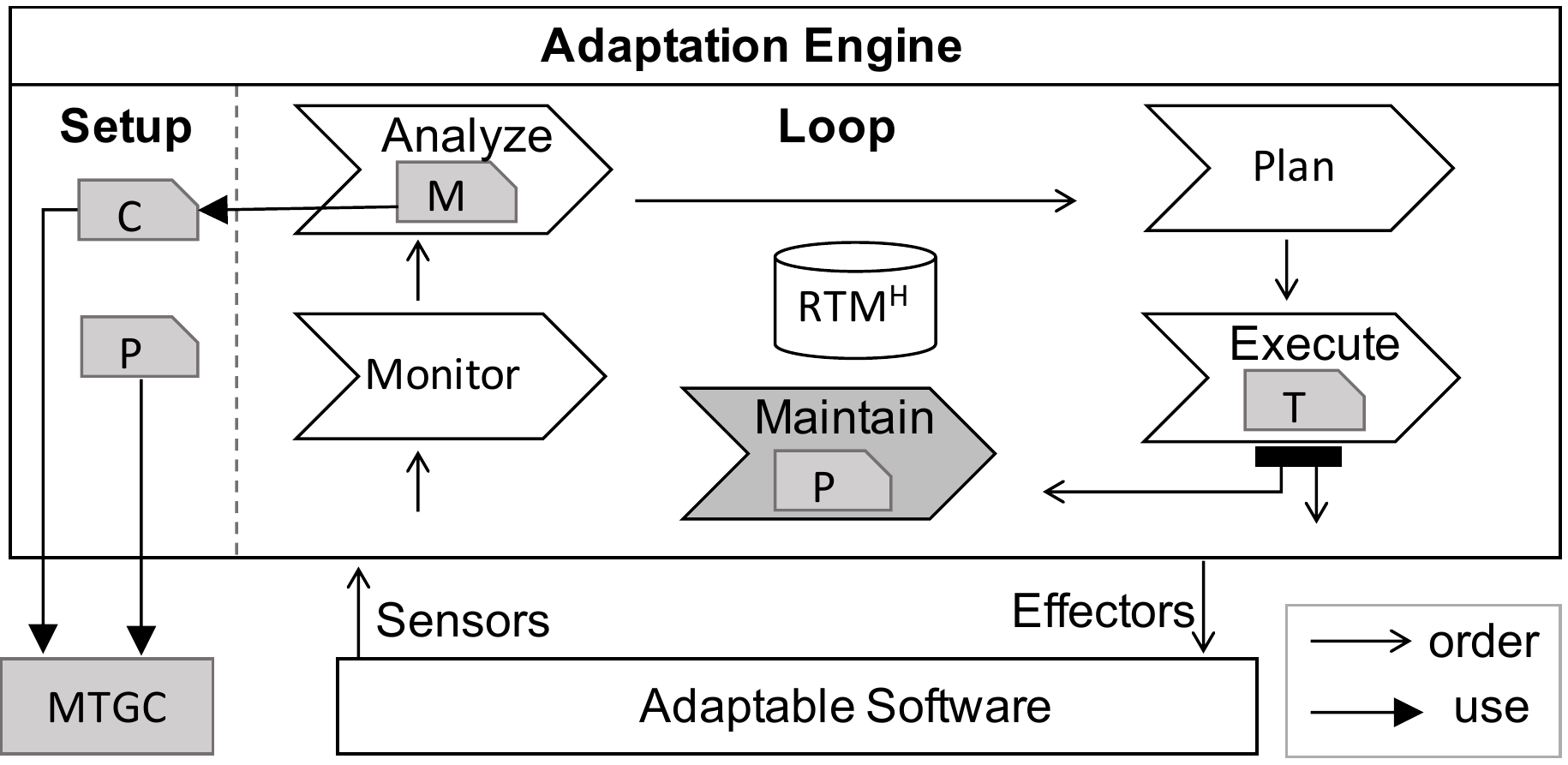}
	\caption{\label{fig:overview}
	Overview of Adaptation Engine}
\end{figure}
In the following, we present the basic modules of our querying scheme named \APPROACHNAME (from \emph{In}cremental queries with \emph{Tempo}ral requirements) together with the extensions that are integrated in an  
\emph{adaptation engine} (\autoref{fig:overview}) to realize 
self-adaptation based on the case-study. The engine consists of the standard MAPE activities, plus the novel and optional \emph{maintain} activity (in gray), sensing and affecting an adaptable software via an \RTMwH.

\subsection{Overview of \APPROACHNAME for Adaptation}
\APPROACHNAME consists of three basic modules. The \emph{C} module (for construction) is executed prior to the adaptation \emph{loop} (during \emph{setup} of the engine) and takes a graph query with temporal requirements captured by an \MTGC as input and constructs a \emph{temporal GDN} by decomposing the query into simpler sub-queries. The module extends the GDN construction presented in \autoref{subsec:query-operationalization} by introducing concepts 
for capturing and handling structural matches whose validity is based on the creation and deletion time point of their elements.
The \emph{M} module (for matching) operates within the adaptation loop (cf.~\autoref{fig:overview}) and executes the temporal GDN, i.e.,~it searches for matches of queries. The module ensures that only (sub-)queries whose matches are affected by changes are re-executed and takes into account temporal requirements on matches. This module is executed in the analysis activity of the loop. 
The \emph{P} module (for pruning) is executed both in setup and within the loop. In setup, it analyzes the query to derive an upper bound on the time window within which elements of the \RTMwH could be used in the evaluation of the query. Then, during the maintain activity of the loop, the module uses these derived \emph{cut-off points} to decide whether to \emph{prune} elements from the \RTMwH that have been removed from the system and are not usable by future query executions. 

The previously presented modules form a (stand-alone) scalable querying scheme which enables the incremental matching of patterns with temporal requirements. The \emph{T} module (for transformation) is an extension that enables self-adaptation via realizing the execution activity of the loop. It processes the query matches, which in this context represent issues requiring adaptation, and performs in-place model transformations, i.e., adaptation actions.

\subsection{Case-Study: Smart Healthcare System}\label{subsec:running-exampl}
\begin{figure}
	\centering
	\includegraphics[width=\linewidth]{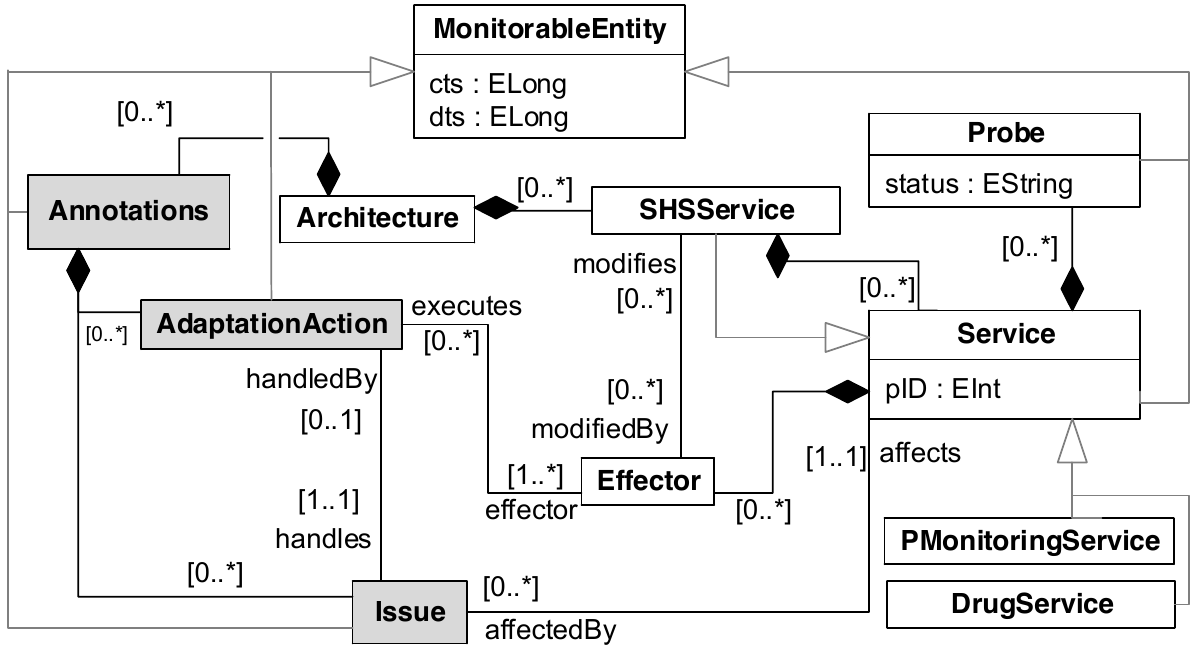}
	\caption{\label{fig:approach-metamodel}
		Metamodel of the SHS (excerpt)}
\end{figure}

Our case-study is based on a service-based simulated Smart Healthcare System (SHS). The SHS is based on smart medical environments \cite{roy2017monitoring} where sensors periodically collect physiological measurements of patients, i.e., data such as temperature, heart-beat, and blood pressure, and certain medical procedures are automated and performed by devices, such as a smart pump administering medicine, based on the collected patient measurements---as otherwise a clinician would be doing. The metamodel of the SHS  (\autoref{fig:approach-metamodel}) is influenced by the exemplar of a self-adaptive service-based medical system in~\cite{DBLP:conf/icse/WeynsC15} and captures an \RTMwH as an instance of the \emph{Architecture} class. To meet the requirements for an \RTMwH, all other elements elements inherit from \emph{MonitorableEntity}, i.e., are equipped with a creation and deletion timestamp.

In our SHS, services are invoked by a main service called \emph{SHSService} to collect measurements (from patient sensors) or take medical actions (via patient effectors, e.g. pump), the former called \emph{PMonitoringService} and the latter \emph{DrugService}. Invocations are triggered by effectors (\emph{Effector}) and invocation results are tracked via monitoring probes (\emph{Probe}) that are attached on \emph{Service}s. \emph{Probe}s are generated periodically or upon events in the real world. Each \emph{Probe} has a \emph{status} attribute whose value depends on the type of \emph{Service}. Each \emph{Service} has a \emph{patientID} attribute which identifies the patient for whom the \emph{Service} is invoked.
\subsection{ History-Aware Self-Adaptation}\label{subsec:MAPE}
In the following, we build on our SHS to envisage a (self-)adaptation scenario that enacts a medical instruction. The instruction imposes temporal requirements on the operation of the SHS which are checked and enforced by the five activities of the adaptation loop which are described below.
The scenario is based on the medical guideline on the treatment of sepsis~\cite{rhodes2017surviving}, a possibly life-threatening condition. We focus on the basic instruction that reads:  ``between \emph{ER Sepsis Triage} and \emph{IV Antibiotics} should be less than 1 hour'', where \emph{ER Sepsis Triage} and \emph{IV Antibiotics} are procedure actions for sepsis documented in real records of patients in a hospital~\cite{DBLP:conf/emisa/MannhardtB17}. Based on the SHS metamodel and the available hospital data, we envisage the procedure described in the guideline performed by the SHS. 

\begin{figure}[t]
	\centering
	\includegraphics[width=\linewidth]{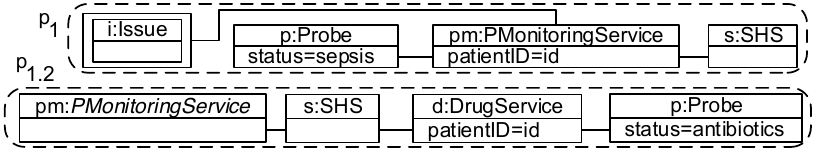}
	\caption{\label{fig:patterns}
		Illustrated Patterns for Case-study}
\end{figure}

In detail, an \emph{ER Sepsis Triage} event is simulated as a \emph{Probe} with a status equal to \emph{sepsis}, generated for a \emph{PMonitoringService pm} which has been invoked by a \emph{SHSService s}. An \emph{IV Antibiotics} event is simulated as a \emph{Probe} with status \emph{antibiotics} from a \emph{Drug Service d} which has also been invoked by \emph{s}. To make sure these two actions are referring to the same patient, we require that the \emph{patientID} of \emph{d} and \emph{pm} are equal. The pattern fragments capturing the occurrence of these events in our SHS are depicted in patterns $p_{1}$ and $p_{1.2}$ in~\autoref{fig:patterns}.
Based on $p_1$ and $p_{1.2}$, the instruction is formulated in \MTGL by the \MTGC $\psi \coloneqq (p_1, (\lozenge_{[0,3600]} \, p_{1.2}))$, that is, for every match of $p_1$ which identifies a (previously untreated) patient with sepsis, eventually in the next hour there is a match for pattern $p_{1.2}$ which identifies the administration of antibiotics to that patient. The system is assumed to track time in seconds. We describe the adaptation activities in detail.

\subsubsection*{Monitor}
During the monitoring activity, the recent events (new readings captured by \emph{Probe}s since the last invocation of the loop) together with their \emph{cts} and \emph{dts} values are reflected in the \RTMwH, which is an instantiation of the \emph{Architecture}. Therefore, the \RTMwH is updated to represent the current architectural system configuration enriched with the relevant temporal data. 

\subsubsection*{Analyze}
The analysis activity detects the potential adaptation issues, which in this context are captured by violations of $\psi$, i.e., the existence in the \RTMwH of structural patterns that reflect sepsis cases ($p_1$) without associated antibiotics ($p_{1.2}$) within one hour. The detection is based on the execution of the temporal GDN by the \emph{M} module of \APPROACHNAME. The temporal GDN is obtained by the \emph{C} module during the setup of the adaptation loop. 

We remark that we aim to detect violations of $\psi$. Therefore, in order for matches to constitute violations of sepsis cases that can be adapted, we execute the following graph query: $q(p_1, \phi_1)$, where $\phi_1 \coloneqq p_1, \neg(\lozenge_{[0,3600]} \, p_{1.2})$. Furthermore, in order to challenge our scheme with a more complicated scenario, we also search for violations for a variation of $\psi$. Namely, that no patient with sepsis should be released from the medical environment prior to being treated, a requirement that resembles conformance checks of medical procedures (cf. \cite{DBLP:conf/emisa/MannhardtB17}). Once more, we rely on the real hospital data and specifically the \emph{Release} event. The structural pattern corresponding to \emph{Release} is pattern $p_{1.1}$, a minor modification of pattern $p_{1.2}$, where the probe attached to a monitoring service has the value \emph{release}. The requirement is captured by $\phi_2 \coloneqq p_1, \neg(\neg p_{1.1} \,\mathrm{U}_{[0,3600]} \, p_{1.2})$. 
Note that an important aspect of analysis is the handling of \emph{potential violations}, i.e., the matching of a sepsis pattern $p_1$ that is not yet associated to an antibiotics pattern $p_{1.2}$ at the current \RTMwH although there is still time for the requirement to be fulfilled in the future. The planning activity only detects these cases as violations if the time difference between the lower bound of the validity interval of the match and the current time point is greater than the until interval in the \MTGC, which, in this case, is 3600 seconds.

The matches detected by the \emph{M} module constitute adaptation issues, and similar to ~\cite{SG-TAAS}, adaptation-related classes (in gray in~\autoref{fig:approach-metamodel}) are employed to facilitate the adaptation. During analysis, the monitoring service involved in detected issues is annotated with an instance of the \emph{Issue} class. Therefore, to ensure that only new violations are matched, $p_1$ contains an \emph{Issue} node \emph{i}  (\autoref{fig:patterns}) surrounded by a box which designates a negative pattern that should not be matched. Issue nodes and other adaptation-related classes are created by ordinary transformation rules.

\subsubsection*{Plan and Execute}
In planning, the engine searches for sepsis probes annotated with an issue. Upon finding them, it attaches an \emph{Effector} on the service to which the probe is attached. In execution, the \emph{T} module searches for effectors and upon finding them takes an adaptation action, i.e., administer antibiotics to the patient via a drug service. This adaptation action is also reflected in the \RTMwH by creating an \emph{AdaptationAction} which is associated to the handled \emph{Issue}. Note that the adaptation has to respect the encoding of the \RTMwH, which means setting the \emph{cts} of created elements appropriately.

\subsubsection*{Maintain}
During maintenance, the \emph{P} module uses the cut-off points derived after the analysis of the \MTGC during setup and \emph{prunes} the \RTMwH, i.e., it removes all elements that have a \emph{dts} in the past and cannot be used in future query evaluations. Following the removal of elements, the GDN is re-executed to update matches.
 
\section{Incremental Matching of Patterns with Temporal Requirements }\label{sec:approach}

In this section we present the inner workings of the matching module of \APPROACHNAME. The module  incrementally searches for matches of a query whose \ac is formulated in \MTGL in a \RTMwH. The matching relies on a decomposition of an \MTGC into simpler queries based on a temporal GDN and the subsequent incremental, bottom-up execution of the latter. 

\subsection{Matches and Their Lifespan}

\begin{figure}
	\includegraphics[width=\linewidth]{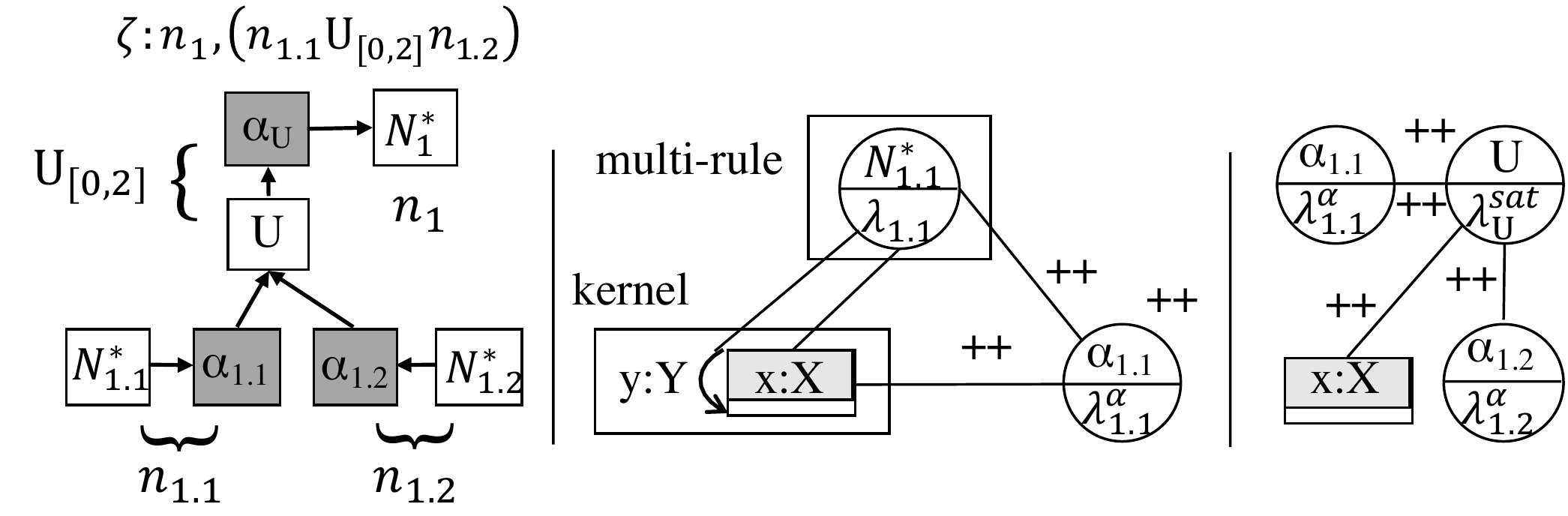}
	\caption{\label{fig:temporal-gdn}
		Temporal GDN: Network and Marking Rules for $\zeta$}
\end{figure}

In the following, we refer to the framework in \cite{DBLP:conf/gg/BeyhlBGL16} (see \autoref{subsec:query-operationalization}) as  \emph{base approach}. The base approach builds on graph queries where application conditions are formulated as \NGCs but does not support the temporal operators \emph{until} and \emph{since} of \MTGL. Building on the base approach, we present our extensions, collectively referred to as \emph{temporal approach} or temporal GDN, which allows for the encoding of temporal operators of \MTGL by a GDN, and thus enable the incremental matching of patterns with temporal requirements.

The temporal approach differs from the base approach (see \autoref{subsec:query-operationalization}) in two key aspects: First, vertices and edges of a \RTMwH encode data on their creation and deletion (via the \emph{cts} and \emph{dts} attributes respectively), i.e. their \emph{lifetime}, which introduces the notion of a \emph{lifespan} of a match, i.e. the time span for which the lifetimes of all nodes and edges of the match overlap; Secondly, \ac that express temporal requirements for structural patterns (enabled by \MTGL) extend the notion of a (structural) match with a period of \emph{temporal validity}. For instance, in $\zeta$ from \autoref{subsec:queries-over-rtmwh}, a match for $n_1$ is not valid (although present in the graph) unless a match for n$_{1.2}$ is found within the specified interval.
Note that, although in the following we consider  both vertices and edges, our implementation aligns with recent approaches and focuses on matching vertices, as, in the general case, edges can be encoded as vertices.

\subsection{Marking Rules for Matches}

The intersection of two intervals is always an interval, whereas the union of two intervals $i_1$ and $i_2$ can only sometimes be encoded as an interval. In this case, we say that $i_1$ and $i_2$ are \emph{adjacent}, i.e. $\textrm{adjacent}(i_1, i_2) = \exists i_\cup \in \mathscr{I}: i_\cup = i_1 \cup i_2$, where $\mathscr{I}$ is the set of all intervals. To encode unions that result in disjoint intervals, i.e. disconnected sets of time points, we define the \emph{fragmented interval} $\mathbb{I}=\{i \,|\, i \in \mathscr{I} \}$. Note that in the following, if we perform set operations on fragmented intervals, we consider the set of time points encoded by the fragmented interval rather than the intervals in $\mathbb{I}$.
To capture the lifespan of matches, we equip the types of marking nodes in the type graph with an attribute $\lambda$ of type fragmented interval. We extend the marking rules of the base approach to set the attribute $\lambda$ of the created marking nodes to the lifespan of the match $m$, where the latter is defined as the intersection of the lifetimes of the matched elements $E$:
\begin{equation}
\label{eq:simple-match}
\lambda^{m}=\bigcap\limits_{e \in E} d^{e}
\end{equation}
where $d^{e}$ is given by $e.\lambda$ if $e$ is a marking node, and by $[e.$\emph{cts}$, e.$\emph{dts}$]$ otherwise. The functionality of the rule remains otherwise unchanged. We name this extended marking rule \MRstar.
\subsection{Marking Rules for Aggregating Matches}
The requirements of \emph{until} in $\zeta$ from the example in \autoref{subsec:queries-over-rtmwh} stipulate that at least a single match for $n_{1.1}$ is present in the graph until at least one $n_{1.2}$ is matched. In order to evaluate this, we need to keep track of the lifespans of all matches for $n_{1.1}$ and $n_{1.2}$. The number of these matches might vary in every GDN update, a property which is not covered by conventional graph transformation rules as presented in \autoref{subsec:graph-based-models}. To allow for a marking node to possibly be associated with a varying number of graph elements, as required by \emph{until} and \emph{since}, we introduce the concept of an \emph{amalgamated marking rule} (\AMR). The latter stems from amalgamated graph transformations \cite{DBLP:conf/birthday/BiermannEEGT10}, where an arbitrary number of parallel transformations are amalgamated, i.e., merged, into a single rule applied to the same model in one transformation step. The \LHS of \AMR contains a \emph{kernel} of graph elements that are bound by the enclosing operator (in $\zeta$, that would be a vertex of type X) and a \emph{multi-rule} which matches an arbitrary number of instances of a certain marking node type. An \AMR thus groups the marking nodes matched by the multi-rule by matches of the kernel. Hence, the \AMR corresponds to a GDN node with a single dependency to the node that creates the marking nodes which the \AMR groups.

Similar to \MRstar, the \RHS of an \AMR creates an $\alpha$ marking node which is connected to the marking nodes of its dependency (matched by the multi-rule) and the elements of its pattern in the kernel marked by those marking nodes. If the dependency is positive, the lifespan of the marking node \AMR is computed by intersecting the lifespan of the match of the kernel $m_{K}$ with the union of the lifespans of the marking nodes $E^{M}$ matched by the multi-rule:
\begin{equation}
\label{eq:alpha}
\lambda^\alpha_{PAC} = \lambda^{m_{K}} \cap \bigcup\limits_{e \in E^{M}} \lambda^{e}
\end{equation}
If the dependency is negative, the relative complement is computed instead of the intersection:
\begin{equation}
\lambda^\alpha_{NAC} = \lambda^{m_{K}} \setminus \bigcup\limits_{e \in E^{M}} \lambda^{e}
\end{equation}
See~\autoref{fig:temporal-gdn} (middle) for an example of \AMR for $n_{1.1}$ from $\zeta$. Note that the depiction of marking rules for the temporal \GDN is illustrated by a split marking node: The bottom compartment contains the lifespan of the marking node.

\subsection{Marking Rules for Temporal Operators}

Finally, we introduce two dedicated marking rules for \emph{until} and \emph{since}, named UMR and SMR respectively, whose \LHS patterns contain elements that are bound by its enclosing query. The UMR and SMR have a dedicated dependency for both their left and right operand and perform a special computation for the lifespan of the marking nodes they create. Let $\lambda^\alpha_\ell$, $\lambda^\alpha_r$ be the lifespan of left, respectively right, operand of \emph{until}. An example of the UMR for $\zeta$  is shown in \autoref{fig:temporal-gdn} (right). The computation of the lifespan of UMR is the following: for every right interval $i \in \lambda^\alpha_r$ and every adjacent left interval $j \in J_i$, where $J_i = \{j \,|\, j \in \lambda^\alpha_\ell \wedge \textrm{adjacent}(i, j)\}$, we compute the \emph{right pivot} interval $\nu$ of $i$ and the \emph{left pivot} interval $\mu$ of $j$.
\begin{equation}
\label{eq:pivot}
\nu(i)=[\ell(i)-u(I_\mathrm{U}), u(i)-\ell(I_\mathrm{U})]
\end{equation}
\begin{equation}
\label{eq:pivot_left}
\mu(j)=[\ell(j), u(j)-\ell(I_\mathrm{U})]
\end{equation}
Considering only adjacent intervals ensures the satisfaction of the requirement of \emph{until} that there is a match of the left operand continuously until there is a match of the right operand. The pivot intervals allow us to check whether the matches of the left and right operand occur with appropriate timing with respect to the relative interval $I_\mathrm{U}$ defined by the \emph{until} operator in the formula, i.e. the interval $[0,2]$ in $\zeta$. The intersection of $\nu(i)$ and $\mu(j)$ then marks an interval where \emph{until} is satisfied. The total satisfaction $\lambda_\mathrm{U}^{sat}$ is computed as the union of all satisfaction intervals:
\begin{equation}
\label{eq:satisfaction}
\lambda_\mathrm{U}^{sat}=\bigcup\limits_{i \in \lambda^\alpha_r}\bigcup\limits_{j \in J_i} \nu(i) \cap \mu(j)
\end{equation}

The computation of the lifespan for a SMR is similar to a UMR. They only differ in the computation of $\nu$ and $\mu$ which for \emph{since} are:
\begin{equation}
\nu(i)=[\ell(i)+\ell(I_\mathrm{S}), u(i)+ u(I_\mathrm{S})]
\end{equation}
\begin{equation}
\mu(j)=[\ell(j)+\ell(I_\mathrm{S}), u(j)]
\end{equation}
where $I_S$ is the specific interval of the \emph{since} operator. For the operators \emph{eventually} and \emph{once}, where the left-hand operand is always true, their $\lambda^\alpha_\ell$ is equal to $\rm I\!R_{0}^{+}$.

The case where $\ell(I_\mathrm{U}) = 0$ (or, in the case of \emph{since}, $\ell(I_\mathrm{S}) = 0$) is special, because according to the specification of \MTGL, the formula can be satisfied without any occurrence of a match for the left operand. Therefore, the computation of $\lambda_\mathrm{U}^{sat}$ is slightly adapted:

\begin{equation}
\label{eq:satisfaction_special_case}
\lambda_\mathrm{U_0}^{sat}=\lambda_\mathrm{U}^{sat} \cup \lambda^\alpha_r
\end{equation}
The computation is analogously adapted for \emph{since}.
\subsection{GDN Construction and Example}
In the previous section, we were concerned with marking nodes created by the GDN rules. In this section, we focus on \GDN nodes which form the components of the network and represent rule applications. Regarding the construction of a temporal GDN from an \MTGC, there are two extensions to the base approach: First, we represent UMR and SMR with two new types of GDN nodes for \emph{until} and \emph{since} respectively. For each such node, we add dependencies to the GDN nodes realizing the left and right operands of the corresponding temporal operator. Secondly, instead of adding direct dependencies from a GDN node $\beta$ to another GDN node $\gamma$, where both $\beta$ and $\gamma$ realize a part of the \MTGC, we construct an intermediate $\alpha$ node, i.e., a GDN node realized by an \AMR. We add a dependency from $\alpha$ to $\gamma$, which is negative iff the original dependency is negative, and a positive dependency from $\beta$ to $\alpha$. Note that this means that in a temporal GDN, only $\alpha$ nodes may have negative dependencies.
\begin{figure}
	\includegraphics[width=0.9\linewidth]{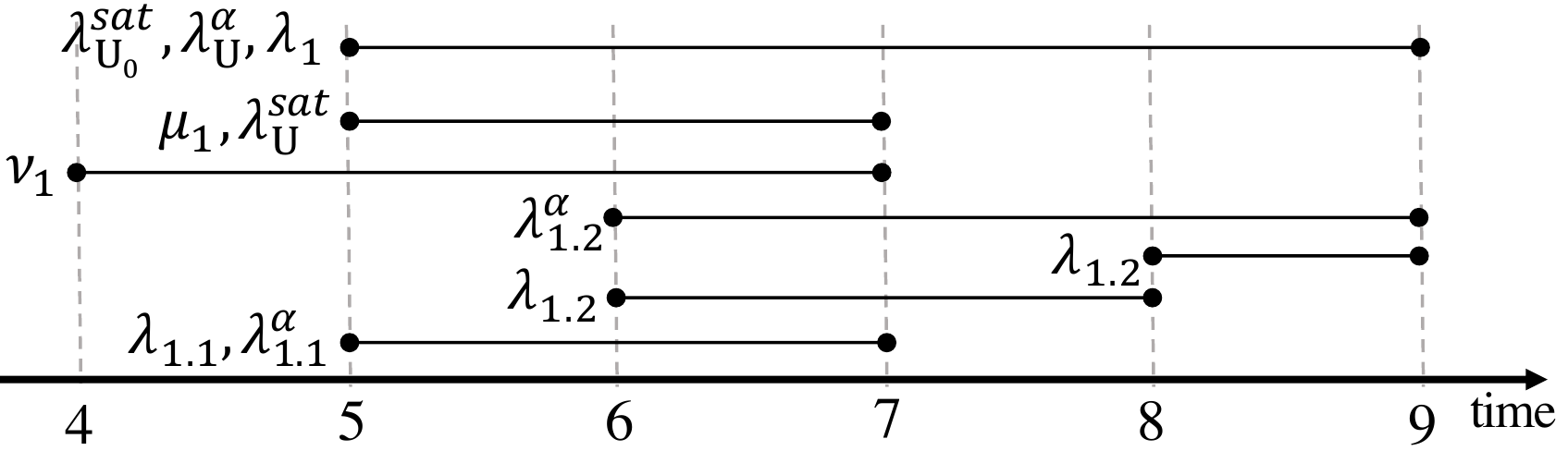}
	\caption{\label{fig:computed-intervals}
		Computed Intervals for $\zeta$ over \RTMwH $H$}
\end{figure}

See \autoref{fig:temporal-gdn} (left) for the temporal GDN of $\zeta$, where the novel $\alpha$ nodes are in dark gray. The patterns $n_{1.1}$ and $n_{1.2}$ are represented by  nodes $N^*_{1.1}$ and $N^*_{1.1}$ respectively. These nodes are both dependencies of their respective $\alpha$ nodes. The $\mathrm{U}$ node is dependent on the two $\alpha$ nodes. Finally, node $N^*_{1}$, the one created for $n_1$, is dependent on an $\alpha$ node with dependency $\mathrm{U}$.
For more complex constructions, the instructions in \cite{DBLP:conf/gg/BeyhlBGL16} apply.
We now present an example of an execution of the query $q(n_1, \zeta)$ over the \RTMwH \emph{H} in \autoref{fig:basic}. In the example, we assume the query is executed at time point $t=9$. The execution consists of seven phases. The computed intervals are illustrated in \autoref{fig:computed-intervals}.
\begin{enumerate}[I]
\item An \MRstar rule for $n_{1.1}$ is applied (node $N^*_{1.1}$ in the \GDN in \autoref{fig:temporal-gdn}). One match is found and thus one marking node is created with $\lambda_{1.1}=[5,7]$, i.e., the lifespan of the match, based on \autoref{eq:simple-match}.
\item The \AMR of node $\alpha_{1.1}$ is applied. One vertex of type X is matched and one marking node of type $N^*_{1.1}$. The marking node created by \AMR groups the lifespans of its dependencies (in this case, only one equal to $[5,7]$) based on \autoref{eq:alpha} and stores the result into its attribute $\lambda_{1.1}^\alpha$.
\item The \MRstar of node $N^*_{1.2}$ is applied. Two matches are found and two marking nodes are created with lifespans $[6,8]$ and $[8,9]$ respectively. 
\item The \AMR of node $\alpha_{1.2}$ is applied. One match of type X is found and two marking nodes of type $N^*_{1.2}$. One marking node is created whose result groups the lifespans of its dependencies and stores it in $\lambda_{1.2}^\alpha=[6,9]$.
\item The UMR of node $\mathrm{U}$ is applied. One match for a vertex of type X is found and two marking nodes for the left and right operand respectively. The lifespans of the right marking node $\lambda_{1.2}^\alpha$ are checked on whether they are adjacent to any of the intervals in the lifespan of the left marking node $\lambda_{1.1}^\alpha$. This is true for two lifespans. The right pivot $\nu_1=[4,7]$ and left pivot $\mu_1=[5,7]$ for these two lifespans are computed based on \autoref{eq:pivot} and \autoref{eq:pivot_left} respectively. The lower bound of \emph{until} is $0$ so we have the special case where the satisfaction interval of \emph{until} is computed by \autoref{eq:satisfaction_special_case} as the union between the intersections of $\mu$ and $\nu$ and the $\lambda_{1.2}^\alpha$ which is $[5,9]$. 
\item The \AMR of node $\alpha_\mathrm{U}$ is applied. The lifespan of the created marking node is computed to be 
$\lambda^\alpha_\mathrm{U}=[5,9]$.
\item The \MRstar of node $N^*_1$ is applied. One vertex of type X is matched and one marking node of type $\alpha_\mathrm{U}$. The lifespan of the match is their intersection: $[5,9]$.
\end{enumerate}
Finally, the executed query returns a match for $n_1$ and the (possibly fragmented) 
interval during which, besides being structurally present in the graph, the match satisfies the temporal requirements expressed by the \MTGC $\zeta$.
Employed for self-adaptation, the temporal \GDN incrementally computes in each execution step all new matches, i.e., adaptation issues, \emph{as soon as possible} and marks them with a marking node. These marking nodes, together with the graph elements they mark, remain in the \RTMwH in subsequent steps.
\section{Memory-Efficient \RTMwH}\label{sec:approach2}

In this section we present the query analysis of the \emph{P} module that allows for memory-efficient \RTMwH. By deletion timestamps, an \RTMwH retains information about elements that have been possibly removed from the represented system. This wealth in insight comes with a price: the perpetual accumulation of historical data causes the \RTMwH to constantly grow in size. A possible remedy is to utilize external, domain-specific  retention policies of patient records, such as the ones publicly available for national healthcare systems, e.g.,~\cite{nhsRecords}. Based on such policies, a process can perform periodical \emph{garbage-collection}, where obsolete elements are pruned from the model and thus memory is freed.

Although such generic removal policies provide a certain upper bound for memory consumption, cluttering the model with obsolete data may lead to deteriorating performance of the pattern matching as more elements have to be considered. A more fine-grained solution than generic policies is to derive this information by the considered queries and their temporal requirements.

We define a function that computes a cut-off point for elements in the \RTMwH based on an \MTGC $\chi_1$ as follows:
\begin{equation}
\label{eq:cut-off}
    \kappa(\chi_1) =
    \begin{cases}
     t' + \max{(\kappa(\chi_{1.1}), \kappa(\chi_{1.2}))} &\text{if } \chi_1 = \chi_{1.1} \mathrm{U}_{[t, t']} \chi_{1.2}\\
     t' + \max{(\kappa(\chi_{1.1}), \kappa(\chi_{1.2}))} &\text{if } \chi_1 = \chi_{1.1} \mathrm{S}_{[t, t']} \chi_{1.2}\\
     \max{(\kappa(\chi_{1.1}), \kappa(\chi_{1.2}))} &\text{if } \chi_1 = \chi_{1.1} \wedge \chi_{1.2}\\
     \kappa(\chi_{1.1}) &\text{if } \chi_1 = \neg\chi_{1.1}\\
     \kappa(\chi_{1.1}) &\text{if } \chi_1 = \exists(n_1, \chi_{1.1})\\
    0 &\text{if } \chi_1 = \TOP\\
    \end{cases}
\end{equation}

Recall that all \MTGCs reduce to $\TOP$ (\autoref{subsec:queries-over-rtmwh}) but for presentation purposes, their syntax has been simplified. Moreover, here we assume that cut-off points need to be calculated for only one query. In case multiple queries are executed at once over the \RTMwH, \autoref{eq:cut-off} has to be adjusted to calculate the upper bound based on all queries in question.

The cut-off point $\kappa(\chi_1)$ corresponds to an upper bound for the maximum number of time units after which a deleted graph element can still be part of a match that may contribute to the satisfaction of the formula $\chi_1$ at the time of checking it. For each graph element $e$, we can hence derive an upper bound of its \emph{relevance window}, i.e. the window during which an element could be used in checking the \ac of queries, by $t_{max}^e = e.dts + \kappa(\chi_1)$.
By deriving the maximum time point that deleted elements can be reused in checking the \ac, we ensure that no element has been pruned prior to the end of its relevance window, and thus that completeness of query results is not affected.
For the formula $\zeta$ from \autoref{subsec:queries-over-rtmwh}, the derived cut-off is 2, i.e., elements need to be retained in the \RTMwH $H$ for 2 time units after their deletion.

Pruning could reduce the size of the matching search space and thus improve the matching time. On the other hand, due to the re-computation of matches after one or more removals have occurred, pruning could potentially incur an increase in the overall adaptation time. Provided such considerations have been made, this step renders an \RTMwH memory-efficient, in that, it is sustainable regardless of whether external memory-saving measures are present or too coarse, e.g., the ones in healthcare mentioned above. 

\begin{figure}
	\includegraphics[width=\linewidth]{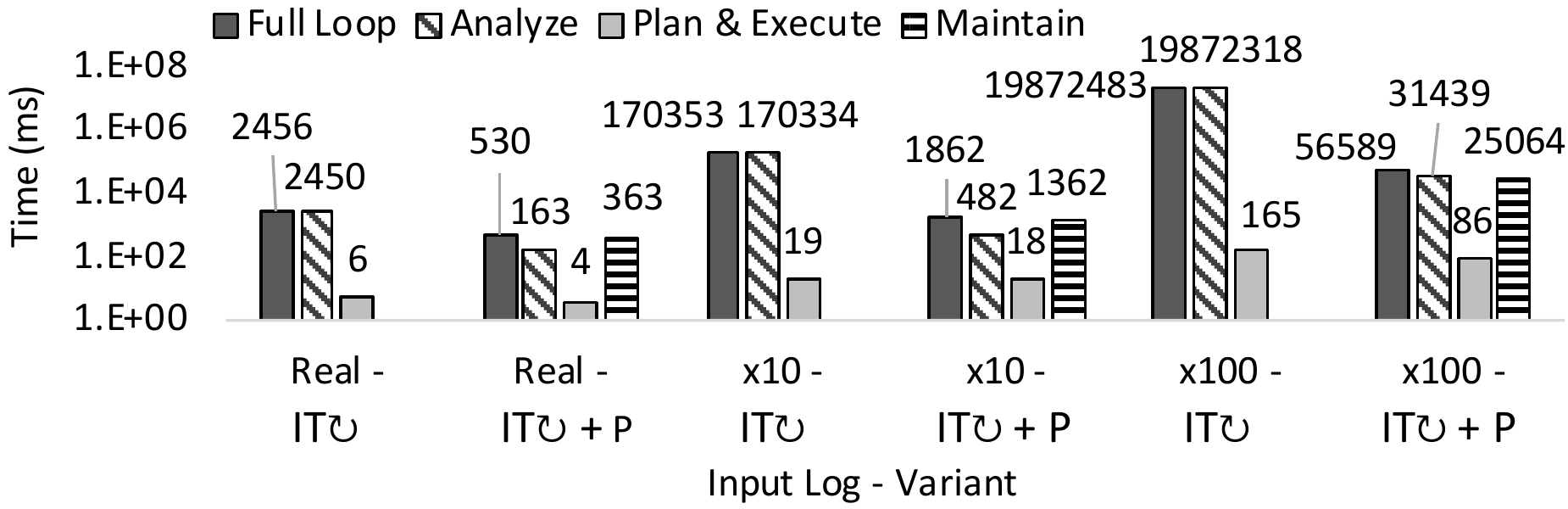}
	\caption{\label{fig:timefi1}
		Cumulative Time of Loop Activities for $\phi_1$}
\end{figure}
\section{Evaluation}
\label{sec:evaluation}
Our implementation of \APPROACHNAME embedded in the adaptation engine (cf.~\autoref{sec:overview})
is based on the Eclipse Modeling Framework (EMF) \cite{emf-website,steinberg2008emf}, which is a widespread MDE technology for creating metamodels of architectures. For pattern matching, we employ the \emph{Story Diagram Interpreter} (SD)~\cite{DBLP:journals/eceasst/GieseHS09} optimized according to~\cite{DBLP:journals/jlap/BarkowskyG20}. SD uses \emph{local search} to start the search from a specific element of the graph and thus reduces the pattern matching effort~\cite{Hildebrandt14}. For computations on intervals we employ an open-source library~\cite{Guava}. For the removal of elements from the runtime model, we transparently replace the native EMF method, via a \textsc{Java} agent, with an optimized version which reduces the potentially expensive shifting of cells in the underlying array list and renders the removal more scalable.

To evaluate our implementation, we developed a simulator of the adaptable SHS presented in~\autoref{subsec:running-exampl}. Our simulations replay events, based on real as well as synthesized patient data on a \RTMwH. The logs are described in~\autoref{subsec:traces}. Based on the processed log event, a corresponding structural fragment is added to the \RTMwH, for instance, an \emph{ER Sepsis Triage} corresponds to the pattern $p_1$ (\autoref{fig:patterns}) being added to the model.
We implemented two variants of   \APPROACHNAME: \textbf{\VARIANTNAME} which contains the construction (\emph{C}), the matching (\emph{M}) and the transformation (\emph{T}) module and  \textbf{\VARIANTNAMEplus} which contains all of the above plus the pruning (\emph{P}) module---the left circle arrow symbolizes the loop-based adaptation. 
 In~\autoref{subsec:scheme+}, we compare the time-scalability of the two variants based on multiple logs with a varying arrival rate of events.\footnote{All experiments and simulations have been conducted on a QuadCore Intel i7 and an OpenJDK8 JVM. Memory measurements are based on values reported by the JVM.}
We compare the performance of \VARIANTNAME and \VARIANTNAMEplus in the detection of adaptation issues, i.e., analysis activity, to a baseline acquired by \textbf{\monpoly}, a state-of-the-art event-based monitoring tool. Besides time- and memory-scalability, the comparison touches on aspects of \emph{usability} and \emph{monitorability}. 
Finally, we discuss threats to validity in~\autoref{subsec:threats}.
\subsection{Input Logs}\label{subsec:traces}
The log used in our experiments~(in the following, \rl) contains $1049$ \emph{trajectories} of sepsis patients admitted to a hospital within $1.5$ years~\cite{DBLP:conf/emisa/MannhardtB17}. Each trajectory comprises a \emph{sequence} of events, among which, we are interested in the \textit{ER Sepsis Triage} (ER), \textit{IV Antibiotics} (IV), and \textit{Release} (RE) events. A trajectory starts with an ER event, and IV and RE events might follow.  
The \emph{inter-arrival time}~(IAT) between two ER events defines the arrival rate of trajectories (as an ER initiates a trajectory).
We use statistical probability distribution fitting to find the best-fitting distribution that characterizes the inter-arrival times between: two ER events~(\trace), an ER and an IV~(\sa), and an ER and an RE~(\sr).
 Then, we use statistical bootstrapping~\cite{diciccio1996bootstrap} to generate two synthetic logs, \emph{x10} and \emph{x100}, with \trace values that are 10 and 100 times smaller respectively than \trace values of the \rl, while \sa and \sr remain as in the \rl. As a result, \emph{x10} and \emph{x100} cover the same period of time as the \rl, and increase the trajectory density (approx.) 10 and 100 times respectively, allowing us to test the scalability of \APPROACHNAME without compromising the statistical characteristics of the \rl.
\subsection{Implementation Variants \VARIANTNAME and \VARIANTNAMEplus}\label{subsec:scheme+}
\begin{figure}[t]
\begin{centering}
  \includegraphics[width=0.95\linewidth]{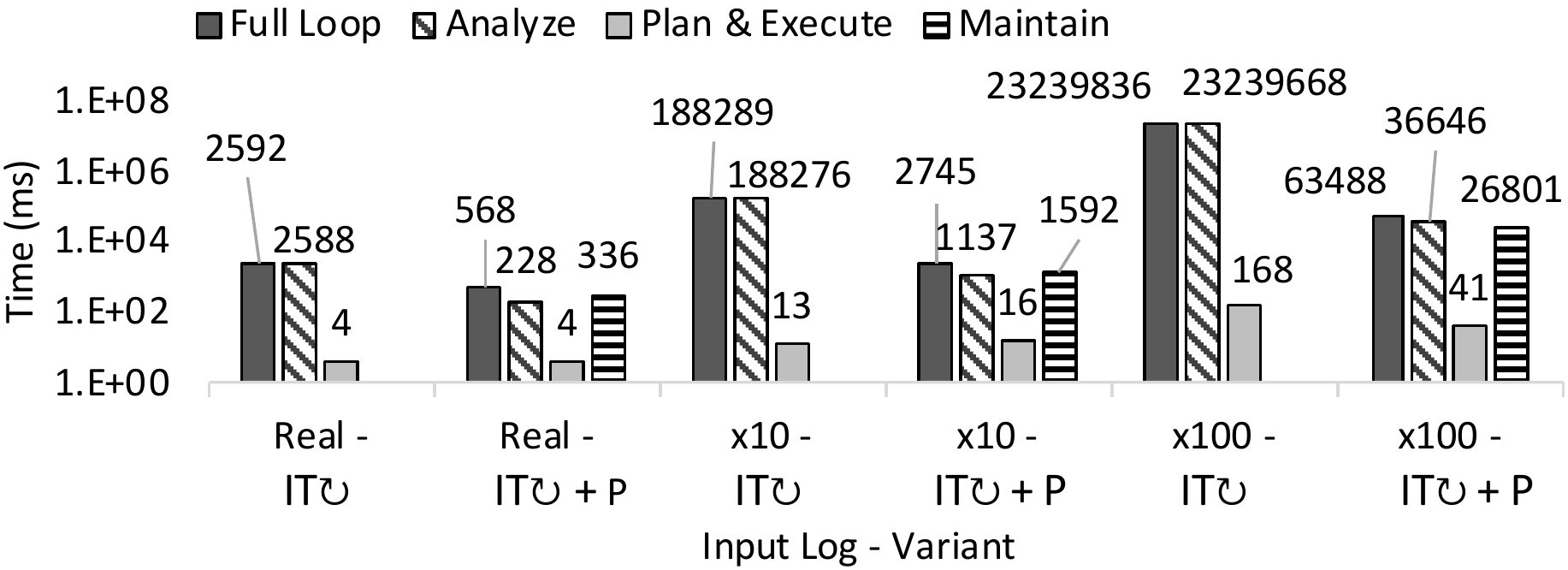}
  \caption{Cumulative Time of Loop Activities for $\phi_2$}
  \label{fig:timefi2}
  \end{centering}
\end{figure}
%
Although pruning the \RTMwH is required for memory-efficiency, we implemented a variant of \APPROACHNAME without the maintain activity, i.e., without pruning. Besides serving as a baseline for \VARIANTNAMEplus,~ \VARIANTNAME could be useful in application domains where it is known that queries of interest change often and thus cut-off points cannot be derived a priori, as historical data might be useful for another query in the future. Moreover, in certain cases, the incurred cost of pruning on the loop execution time might be undesirable.

We evaluate \VARIANTNAME and \VARIANTNAMEplus with respect to their \emph{reaction time} (or loop time) over an increasing amount of log events and model sizes. In this context, the reaction time is equal to the required time for a loop, i.e., the time from when an issue is detected to when a corresponding adaptation action has been performed. Thus the reaction time consists of times for analysis, planning, execution and, for \VARIANTNAMEplus, maintenance time. The time spent in monitoring, i.e., processing an event and adding the corresponding fragment to the \RTMwH, is negligible and thus not measured.
A loop is invoked periodically based on a predefined but modifiable frequency. In our experiments, based on the \trace of the logs, we set the invocation frequency to one hour, to avoid frequent 
invocations where there are no events to be processed.
The invocation frequency coincides with the maximum delay of a violation detection, i.e., in the worst case, a violation will occur just after the loop and will be detected at the next invocation which in this case is exactly after one hour.

The \emph{C} module of \APPROACHNAME as well as the function $\kappa$ (\autoref{sec:approach2}), that derives the cut-off points used by \emph{P} module of \VARIANTNAMEplus, are executed only once during the setup of the loop. Each experiment proceeds as follows: Events from the logs are processed and changes are made to the \RTMwH; The loop is invoked at the predefined intervals which includes the analysis activity, where the \emph{M} module executes queries that search for violations of $\phi_1$ and $\phi_2$ (see \autoref{subsec:running-exampl}) that constitute adaptation issues, and the \emph{T} module performs transformations corresponding to adaptation actions; Then, for \VARIANTNAMEplus, maintenance is done and matches are recomputed.

\begin{figure}[t]
\begin{centering}
  \includegraphics[width=0.9\linewidth]{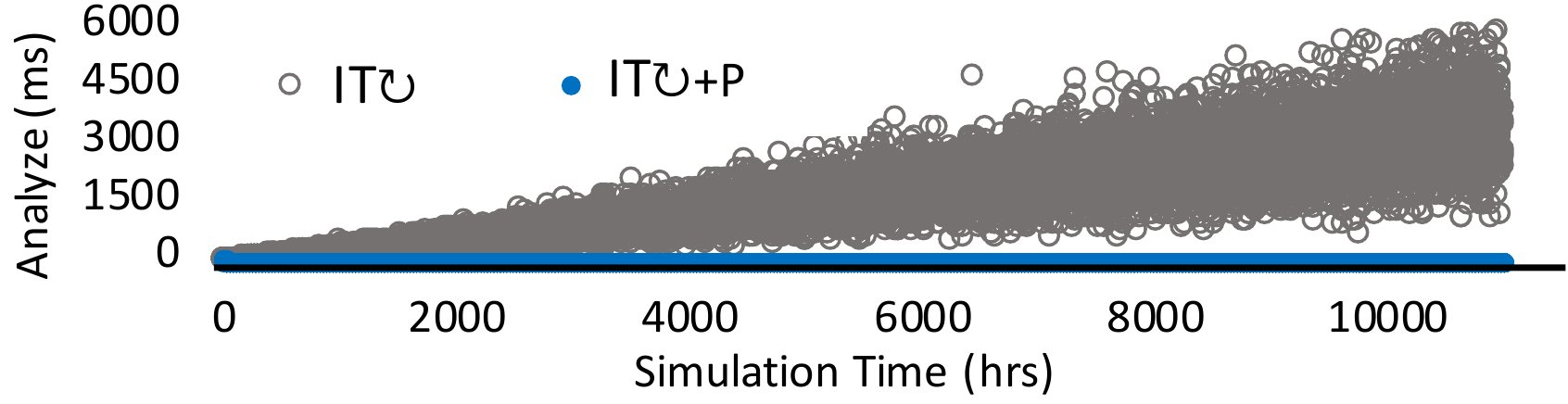}
  \caption{Analysis Time for Engine Variants 
  ($\phi_1$ - x100)}
  \label{fig:analysisPhi1X100}
  \end{centering}
\end{figure}

The experiments simulate the data in \emph{real}, $x10$, and $x100$ logs. Each experiment entails the execution of one variant measured for one performance aspect (time or memory). 
\autoref{fig:timefi1} and \autoref{fig:timefi2} depict the cumulative time (in logarithmic scale) for each of the measured loop activities 
and the reaction, i.e. total, time for $\phi_1$ and $\phi_2$.
As expected, the results are mainly influenced by the analysis activity, which is when issues are detected. Two parameters in the conducted experiments increase simultaneously: first, the number of processed events (total number and per loop) and, second, the size of the \RTMwH over which the queries are executed. The 
analysis time of \VARIANTNAME increases with respect to these two parameters but at a smaller pace. Thanks to pruning, \VARIANTNAMEplus's 
analysis time increases yet at a considerably smaller pace compared to \VARIANTNAME. However, since pruning forces a re-computation of the results, 
the time it requires is non-negligible. The analysis time for each loop of the two variants for the \emph{x100} log is shown in \autoref{fig:analysisPhi1X100}. The pruning of \RTMwH allows the analysis time of \VARIANTNAMEplus to remain constant. 

\subsection{Comparison to State-of-the-art Tool}
\label{subsec:monpoly}

\begin{table}[b]\centering
 	\caption {Memory Consumption (max) for $\phi_1$ (MB)}
	\label{tab:mem}
 	\begin{tabular}{l r r r}\hline
 		& \emph{real} &\emph{x10}&\emph{x100}\\
 		\cmidrule{2-4}
\VARIANTNAME& 43 & 174	&1544\\
 	\VARIANTNAMEplus & 29  & 30 &33
 \\
 	\monpoly&20&	31&	165\\ \hline
 	\end{tabular} 
\end{table}
During the analysis activity, \APPROACHNAME processes a sequence of events which represents an ongoing system execution and checks whether the observed sequence (captured in the \RTMwH) satisfies a formal specification (captured by one or more \MTGCs). This monitoring approach is also known as Runtime Monitoring (RM)~\cite{DBLP:series/lncs/BartocciDDFMNS18} and we therefore employ a prominent RM tool in order to acquire a baseline for the performance of \VARIANTNAME and \VARIANTNAMEplus in detecting issues during analysis. We compare to  \monpoly~\cite{DBLP:conf/rv/BasinKZ17,Basin_2015_MonitoringMetricFirstOrderTemporalProperties}, a mature command-line tool which notably combines an adequately expressive specification language with an efficient incremental monitoring algorithm and has been the reference point in evaluations of  other RM tools ~\cite{Havelund_2018_EfficientRuntimeVerificationofFirstOrderTemporalProperties, Dou_2017_AModelDrivenApproachtoTraceCheckingofPatternBasedTemporalProperties} and among top-performers in an RM competition~\cite{DBLP:journals/sttt/BartocciFBCDHJK19}. Its specification language is the Metric First-Order Temporal Logic \cite{Basin_2015_MonitoringMetricFirstOrderTemporalProperties} (MFOTL) which employs first-order \emph{relations} to capture system entities and their relationships. The usage of a temporal logic
facilitates the translation of temporal requirements between \monpoly and our implementation.
For the encoding of the SHS metamodel by relations we translate each edge following standard practices (cf.~\cite{Rensink_2004_Representingfirstorderlogicusinggraphs}) and each vertex into a relation with an arity that depends on the number of class attributes. Based on this encoding, we translate each log event into a series of relations.

Encoding a graph pattern in MFOTL requires an explicit definition of the expected temporal ordering of the events that corresponds to the order of creation of the elements in the simulation. To emulate pattern matching, we would therefore have to build an MFOTL formula that would consider all possible events as a start for matching the pattern and then search in the past of the execution or in the present for the rest of the events.\footnote{Since \monpoly outputs the time point of a violation, forward-looking matching, i.e., matching a relation in the past and subsequently searching for other relations in its future, would not produce the desired result as it would always only output the time point the first violating relation was matched.} 
Leveraging the knowledge of the actual order in which events occur in the simulation, we simplify the formulas for \monpoly by formulating only the correct ordering. This creates an advantage for \monpoly in the comparison with our implementation. The difficulties in emulating pattern matching with \monpoly indicate the tool is sub-optimal for graph-based models and pattern matching. We map $\phi_1$ in a straightforward manner to its MFOTL equivalent, i.e., the temporal operators remain intact and relation fragments are used instead of patterns. This is not possible for $\phi_2$, as \monpoly restricts the use of negation in this case. It does so for reasons of monitorability, as the tool assumes an infinite domain of values, and the negation of  $p_{1.1}$ at a given time point when it does not exist is satisfied by infinite values and thus non-monitorable. In the following, we compare to \monpoly only for $\phi_1$.

\begin{figure}[t]
\begin{centering}
  \includegraphics[width=0.95\linewidth]{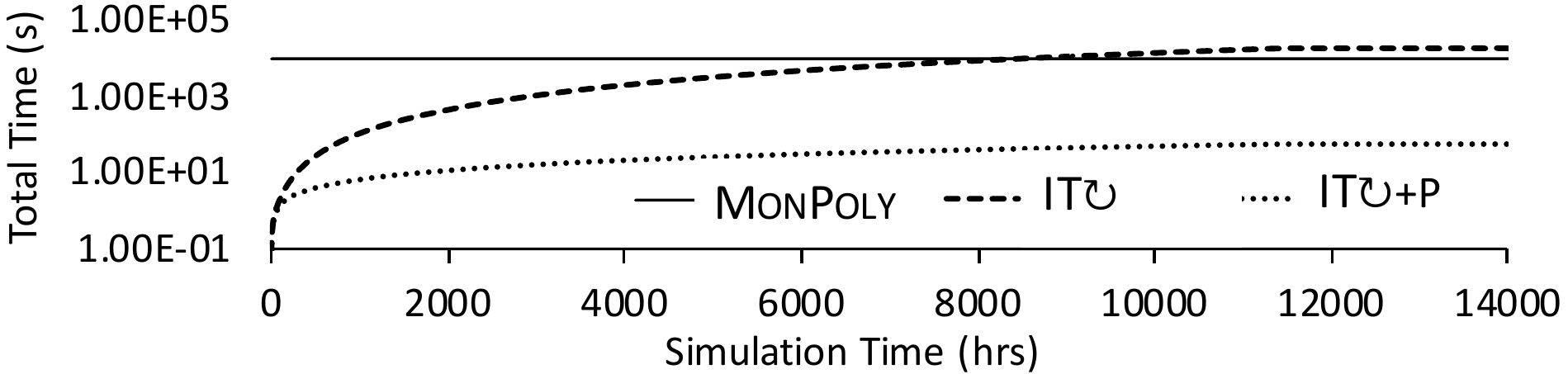}
  \caption{Total Experiment Time vs. \monpoly for  $\phi_1$}
  \label{fig:totaltimefi1}
  \end{centering}
\end{figure}
We acquire the baseline by executing \monpoly only once at the end of each simulation. The latest \monpoly version (1.1.9) was used and run on the same machine as the implementation variants. The results for the execution time (in seconds) for \emph{x100}, which emphasizes the trends of smaller logs, are shown in \autoref{fig:totaltimefi1}.
The results are compared to the \emph{total experiment time} (not only analysis time) of our variants
as other adaptation-related activities could affect their analysis times. Issue detection with \monpoly is faster than \VARIANTNAME, however, \VARIANTNAMEplus, due to pruning, outperforms \monpoly. \autoref{tab:mem} shows similar results for memory consumption.
\subsection{Threats to Validity}\label{subsec:threats}
Threats to internal validity concern the experimental setting. We systematically evaluated \VARIANTNAME, \VARIANTNAMEplus, and \monpoly by using a controlled simulation of an SHS. Our focus was the effects of incremental pattern matching and pruning on the time- and memory-scalability of the variants measured during the analysis activity of the adaptation loop. To solely focus on these effects, the two variants share identical monitoring, planning, and execution activities and they use the same architectural metamodel. \monpoly is evaluated only on the analysis. The experiments draw from an example where instructions are deterministic. Moreover, the experiments simulate multiple input logs with different properties to allow for testing the variants in different circumstances such as increasing system load. The logs used are either real data or data extracted from real data employing sophisticated statistical bootstrapping.

Threats to external validity may restrict the generalization of our evaluation results outside the scope of our experiments. We evaluated our querying scheme on simulated real and synthetic data while enacting an instruction from a real medical guideline \cite{rhodes2017surviving}. Our SHS metamodel is influenced by a peer-reviewed self-adaptation artifact. As a result, we have confidence that our evaluation, to a certain extent, holds for real scenarios. While our experiments can serve as an indication to the scalability of our querying scheme, quantitative claims on scalability require more extensive simulation scenarios taking into consideration real IoT device properties, such as memory.
\monpoly is not built for pattern matching and our emulation of the latter might have room for improvement. \monpoly's semantics is \emph{point-based} while the semantics of \MTGL (and thus of our implementation) is interval-based. This fundamental difference did not impact our experiments but might influence more extensive comparisons. For the property that \monpoly could not monitor, there might exist equivalent, monitorable MFOTL formulas which, however, would not correspond to the \MTGC straightforwardly.
\section{Related Work}\label{sec:related}
The efficient storage of historical data from the system execution in a (graph-based) model has been the focus of extensive research, e.g. temporal graphs~\cite{DBLP:conf/seke/0001FJRT17}, the same is not true however for the scalable querying and the sustainability of runtime models~\cite{DBLP:journals/sosym/BencomoGS19}. Recent works build on a database, either graph-~\cite{Bencomo2019} or map-based~\cite{ DBLP:conf/er/GomezCW18} to store model versions 
which are queried by means of an OCL extension that supports temporal primitives. Neither~\cite{Bencomo2019} nor~\cite{DBLP:conf/er/GomezCW18} consider an online setting 
where query matches can be utilized while the system is running. In an online setting, storing multiple versions of the model and accesses to a database storage take a significant toll on real-time querying performance (the latter indicated by the evaluation in~\cite{DBLP:conf/er/GomezCW18}), especially for far-reaching past queries. To improve performance, the authors in~\cite{Bencomo2019} provide the capability to a priori manually annotate such queries, such that their matches are pre-computed while the system evolves, which, however, is automatically achieved by the temporal GDN of \APPROACHNAME. Moreover, \APPROACHNAME uses an in-memory representation of the model, as shared memory space generally makes the real-time querying faster, which is key for the online setting and the (adaptive) systems of interest in this work. Finally, a solution for the perpetual accumulation of historical data is missing from both~\cite{Bencomo2019} and~\cite{DBLP:conf/er/GomezCW18}. 

The setting of our case-study resembles \emph{streaming}~\cite{DBLP:conf/icmt/CuadradoL13} and \emph{active model transformations}~\cite{DBLP:conf/models/BeaudouxBBJ10} where the model and query results are assumed to be continuously updated by a stream of model elements or events that are mapped to model elements. These paradigms increase demands on the performance of the pattern matching, which previous approaches have met via employing incremental query evaluation frameworks~\cite{DBLP:journals/scp/UjhelyiBHHIRSV15} (similarly to \APPROACHNAME) and the distribution of pattern matching~\cite{DBLP:conf/models/SzarnyasIRHBV14}. The approach in~\cite{David_2014_StreamingModelTransformationsByComplexEventProcessing,David_2018_FoundationsforStreamingModelTransformationsbyComplexEventProcessing} generates events when patterns are matched and then employs \emph{complex event processing} to check whether generated events occur within a given time window, thus capturing, albeit compositely and to a certain extent, temporal requirements on matched patterns. Contrary to these approaches, \APPROACHNAME natively encompasses the history of model evolution in the model representation, the query specification language, as well as during pattern matching.

As previously mentioned, our setting is also related to Runtime Monitoring (see \autoref{subsec:monpoly}). Besides \monpoly however, other approaches provide no or only partial support for key features of \APPROACHNAME such as events containing data, temporal requirements, or metric temporal operators: The work in \cite{Dou_2017_AModelDrivenApproachtoTraceCheckingofPatternBasedTemporalProperties} concerns propositional events, i.e., containing no data, and is thus unsuitable for the use-cases discussed in this paper; In~\cite{Bur_2018_DistributedGraphQueriesforRuntimeMonitoringofCyberPhysicalSystems,DBLP:journals/sttt/BurSVV20} an event-based scheme for the incremental matching of graph patterns in a runtime model is presented which however does not support the integration of model queries and temporal requirements; the tool in ~\cite{DBLP:conf/fmcad/HavelundPU17, Havelund_2018_EfficientRuntimeVerificationofFirstOrderTemporalProperties}  employs relations and discards unusable data (similar to pruning) but its logic supports only past operators without intervals.

In~\cite{2020_towards_highly_scalable_runtime_models_with_history} we presented a preliminary version of \APPROACHNAME that is based on an ad hoc, manual translation of a single, syntactically restricted past \MTGC that does not account for the aggregation of matches and also presents an ad hoc, manual derivation of cut-off points. The \APPROACHNAME version presented here translates \MTGCs and derives cut-off points automatically. Moreover, it introduces the required concepts and facilities for aggregating matches, supports both past and future operators, and enables a complete adaptation loop.
In~\cite{DBLP:conf/gg/SchneiderSMG20} we presented the formal foundation for the translation of a syntactically restricted \MTGC to an \NGC which is then checked against an event-based execution aggregated in an \RTMwH-like graph. However, the approach does not consider model queries, rather a non-incremental  satisfaction check of the \MTGC, nor does it consider past operators or a means to limit data accumulation.
\section{Conclusion and Future Work}\label{sec:conclusion}
We have introduced a querying scheme where graph-based model queries are integrated with temporal requirements on patterns, formulated in a temporal graph logic. Our scheme enables the incremental execution of queries over runtime models with history. Building on self-adaptive systems, in our case-study query matches capture adaptation issues in the runtime model which are handled by in-place model transformations. Our scheme offers the option to retain in the model only information that are relevant to the query executions. We present an implementation which we evaluate based on a simulation of both real as well as synthetic data and compare its efficiency in detecting issues to a relevant monitoring tool.

As future work, we plan to present a formalization of our approach, integrate more sophisticated decision-making schemes in the planning phase, improve the performance of our implementation by employing indexing structures that can index matches based on their intervals, and evaluate the performance of other incremental query evaluation frameworks such as RETE networks.
\begin{acks}
This work is partially supported by the German Research Foundation (DFG) under GI 765/8-1.
\end{acks}
\bibliographystyle{ACM-Reference-Format}
\bibliography{scalable-querying}


\begin{thebibliography}{59}


\ifx \showCODEN    \undefined \def \showCODEN     #1{\unskip}     \fi
\ifx \showDOI      \undefined \def \showDOI       #1{#1}\fi
\ifx \showISBNx    \undefined \def \showISBNx     #1{\unskip}     \fi
\ifx \showISBNxiii \undefined \def \showISBNxiii  #1{\unskip}     \fi
\ifx \showISSN     \undefined \def \showISSN      #1{\unskip}     \fi
\ifx \showLCCN     \undefined \def \showLCCN      #1{\unskip}     \fi
\ifx \shownote     \undefined \def \shownote      #1{#1}          \fi
\ifx \showarticletitle \undefined \def \showarticletitle #1{#1}   \fi
\ifx \showURL      \undefined \def \showURL       {\relax}        \fi
\providecommand\bibfield[2]{#2}
\providecommand\bibinfo[2]{#2}
\providecommand\natexlab[1]{#1}
\providecommand\showeprint[2][]{arXiv:#2}

\bibitem[\protect\citeauthoryear{Artale, Parent, and Spaccapietra}{Artale
  et~al\mbox{.}}{2007}]%
        {artale2007evolving}
\bibfield{author}{\bibinfo{person}{Alessandro Artale},
  \bibinfo{person}{Christine Parent}, {and} \bibinfo{person}{Stefano
  Spaccapietra}.} \bibinfo{year}{2007}\natexlab{}.
\newblock \showarticletitle{Evolving objects in temporal information systems}.
\newblock \bibinfo{journal}{\emph{Annals of Mathematics and Artificial
  Intelligence}} \bibinfo{volume}{50}, \bibinfo{number}{1-2}
  (\bibinfo{year}{2007}), \bibinfo{pages}{5--38}.
\newblock


\bibitem[\protect\citeauthoryear{Barkowsky and Giese}{Barkowsky and
  Giese}{2020}]%
        {DBLP:journals/jlap/BarkowskyG20}
\bibfield{author}{\bibinfo{person}{Matthias Barkowsky} {and}
  \bibinfo{person}{Holger Giese}.} \bibinfo{year}{2020}\natexlab{}.
\newblock \showarticletitle{Hybrid search plan generation for generalized graph
  pattern matching}.
\newblock \bibinfo{journal}{\emph{J. Log. Algebraic Methods Program.}}
  \bibinfo{volume}{114} (\bibinfo{year}{2020}), \bibinfo{pages}{100563}.
\newblock
\urldef\tempurl%
\url{https://doi.org/10.1016/j.jlamp.2020.100563}
\showDOI{\tempurl}


\bibitem[\protect\citeauthoryear{Bartocci, Deshmukh, Donz{\'{e}}, Fainekos,
  Maler, Nickovic, and Sankaranarayanan}{Bartocci et~al\mbox{.}}{2018}]%
        {DBLP:series/lncs/BartocciDDFMNS18}
\bibfield{author}{\bibinfo{person}{Ezio Bartocci},
  \bibinfo{person}{Jyotirmoy~V. Deshmukh}, \bibinfo{person}{Alexandre
  Donz{\'{e}}}, \bibinfo{person}{Georgios~E. Fainekos}, \bibinfo{person}{Oded
  Maler}, \bibinfo{person}{Dejan Nickovic}, {and} \bibinfo{person}{Sriram
  Sankaranarayanan}.} \bibinfo{year}{2018}\natexlab{}.
\newblock \showarticletitle{Specification-Based Monitoring of Cyber-Physical
  Systems: {A} Survey on Theory, Tools and Applications}.
\newblock In \bibinfo{booktitle}{\emph{Lectures on Runtime Verification -
  Introductory and Advanced Topics}}, \bibfield{editor}{\bibinfo{person}{Ezio
  Bartocci} {and} \bibinfo{person}{Yli{\`{e}}s Falcone}} (Eds.).
  \bibinfo{series}{Lecture Notes in Computer Science},
  Vol.~\bibinfo{volume}{10457}. \bibinfo{publisher}{Springer},
  \bibinfo{pages}{135--175}.
\newblock
\urldef\tempurl%
\url{https://doi.org/10.1007/978-3-319-75632-5\_5}
\showDOI{\tempurl}


\bibitem[\protect\citeauthoryear{Bartocci, Falcone, Bonakdarpour, Colombo,
  Decker, Havelund, Joshi, Klaedtke, Milewicz, Reger, Rosu, Signoles, Thoma,
  Zalinescu, and Zhang}{Bartocci et~al\mbox{.}}{2019}]%
        {DBLP:journals/sttt/BartocciFBCDHJK19}
\bibfield{author}{\bibinfo{person}{Ezio Bartocci}, \bibinfo{person}{Yli{\`{e}}s
  Falcone}, \bibinfo{person}{Borzoo Bonakdarpour}, \bibinfo{person}{Christian
  Colombo}, \bibinfo{person}{Normann Decker}, \bibinfo{person}{Klaus Havelund},
  \bibinfo{person}{Yogi Joshi}, \bibinfo{person}{Felix Klaedtke},
  \bibinfo{person}{Reed Milewicz}, \bibinfo{person}{Giles Reger},
  \bibinfo{person}{Grigore Rosu}, \bibinfo{person}{Julien Signoles},
  \bibinfo{person}{Daniel Thoma}, \bibinfo{person}{Eugen Zalinescu}, {and}
  \bibinfo{person}{Yi Zhang}.} \bibinfo{year}{2019}\natexlab{}.
\newblock \showarticletitle{First international Competition on Runtime
  Verification: rules, benchmarks, tools, and final results of {CRV} 2014}.
\newblock \bibinfo{journal}{\emph{{STTT}}} \bibinfo{volume}{21},
  \bibinfo{number}{1} (\bibinfo{year}{2019}), \bibinfo{pages}{31--70}.
\newblock
\urldef\tempurl%
\url{https://doi.org/10.1007/s10009-017-0454-5}
\showDOI{\tempurl}


\bibitem[\protect\citeauthoryear{Basin, Klaedtke, Müller, and
  Zălinescu}{Basin et~al\mbox{.}}{2015}]%
        {Basin_2015_MonitoringMetricFirstOrderTemporalProperties}
\bibfield{author}{\bibinfo{person}{David Basin}, \bibinfo{person}{Felix
  Klaedtke}, \bibinfo{person}{Samuel Müller}, {and} \bibinfo{person}{Eugen
  Zălinescu}.} \bibinfo{year}{2015}\natexlab{}.
\newblock \showarticletitle{Monitoring {Metric} {First}-{Order} {Temporal}
  {Properties}}.
\newblock \bibinfo{journal}{\emph{J. ACM}} \bibinfo{volume}{62},
  \bibinfo{number}{2} (\bibinfo{date}{May} \bibinfo{year}{2015}),
  \bibinfo{pages}{1--45}.
\newblock
\showISSN{00045411}
\urldef\tempurl%
\url{https://doi.org/10.1145/2699444}
\showDOI{\tempurl}


\bibitem[\protect\citeauthoryear{Basin, Klaedtke, and Zalinescu}{Basin
  et~al\mbox{.}}{2017}]%
        {DBLP:conf/rv/BasinKZ17}
\bibfield{author}{\bibinfo{person}{David~A. Basin}, \bibinfo{person}{Felix
  Klaedtke}, {and} \bibinfo{person}{Eugen Zalinescu}.}
  \bibinfo{year}{2017}\natexlab{}.
\newblock \showarticletitle{The MonPoly Monitoring Tool}. In
  \bibinfo{booktitle}{\emph{RV-CuBES 2017. An International Workshop on
  Competitions, Usability, Benchmarks, Evaluation, and Standardisation for
  Runtime Verification Tools, September 15, 2017, Seattle, WA, {USA}}}
  \emph{(\bibinfo{series}{Kalpa Publications in Computing},
  Vol.~\bibinfo{volume}{3})}, \bibfield{editor}{\bibinfo{person}{Giles Reger}
  {and} \bibinfo{person}{Klaus Havelund}} (Eds.).
  \bibinfo{publisher}{EasyChair}, \bibinfo{pages}{19--28}.
\newblock
\urldef\tempurl%
\url{http://www.easychair.org/publications/paper/62MC}
\showURL{%
\tempurl}


\bibitem[\protect\citeauthoryear{Beaudoux, Blouin, Barais, and
  J{\'e}z{\'e}quel}{Beaudoux et~al\mbox{.}}{2010}]%
        {DBLP:conf/models/BeaudouxBBJ10}
\bibfield{author}{\bibinfo{person}{Olivier Beaudoux}, \bibinfo{person}{Arnaud
  Blouin}, \bibinfo{person}{Olivier Barais}, {and} \bibinfo{person}{Jean-Marc
  J{\'e}z{\'e}quel}.} \bibinfo{year}{2010}\natexlab{}.
\newblock \showarticletitle{Active operations on collections}. In
  \bibinfo{booktitle}{\emph{International Conference on Model Driven
  Engineering Languages and Systems}}. Springer, \bibinfo{pages}{91--105}.
\newblock


\bibitem[\protect\citeauthoryear{Bencomo, France, Cheng, and
  A{\ss}mann}{Bencomo et~al\mbox{.}}{2014}]%
        {bencomo2014models}
\bibfield{author}{\bibinfo{person}{Nelly Bencomo}, \bibinfo{person}{Robert~B
  France}, \bibinfo{person}{Betty~HC Cheng}, {and} \bibinfo{person}{Uwe
  A{\ss}mann}.} \bibinfo{year}{2014}\natexlab{}.
\newblock \bibinfo{booktitle}{\emph{Models@ run. time: foundations,
  applications, and roadmaps}}. Vol.~\bibinfo{volume}{8378}.
\newblock \bibinfo{publisher}{Springer}.
\newblock


\bibitem[\protect\citeauthoryear{Bencomo, G{\"o}tz, and Song}{Bencomo
  et~al\mbox{.}}{2019a}]%
        {Bencomo.2019.Models}
\bibfield{author}{\bibinfo{person}{Nelly Bencomo}, \bibinfo{person}{Sebastian
  G{\"o}tz}, {and} \bibinfo{person}{Hui Song}.}
  \bibinfo{year}{2019}\natexlab{a}.
\newblock \showarticletitle{Models@run.time: a guided tour of the state of the
  art and research challenges}.
\newblock \bibinfo{journal}{\emph{Software {\&} Systems Modeling}}
  (\bibinfo{year}{2019}).
\newblock


\bibitem[\protect\citeauthoryear{Bencomo, G{\"{o}}tz, and Song}{Bencomo
  et~al\mbox{.}}{2019b}]%
        {DBLP:journals/sosym/BencomoGS19}
\bibfield{author}{\bibinfo{person}{Nelly Bencomo}, \bibinfo{person}{Sebastian
  G{\"{o}}tz}, {and} \bibinfo{person}{Hui Song}.}
  \bibinfo{year}{2019}\natexlab{b}.
\newblock \showarticletitle{Models@run.time: a guided tour of the state of the
  art and research challenges}.
\newblock \bibinfo{journal}{\emph{Software and Systems Modeling}}
  \bibinfo{volume}{18}, \bibinfo{number}{5} (\bibinfo{year}{2019}),
  \bibinfo{pages}{3049--3082}.
\newblock
\urldef\tempurl%
\url{https://doi.org/10.1007/s10270-018-00712-x}
\showDOI{\tempurl}


\bibitem[\protect\citeauthoryear{Beyhl, Blouin, Giese, and Lambers}{Beyhl
  et~al\mbox{.}}{2016}]%
        {DBLP:conf/gg/BeyhlBGL16}
\bibfield{author}{\bibinfo{person}{Thomas Beyhl}, \bibinfo{person}{Dominique
  Blouin}, \bibinfo{person}{Holger Giese}, {and} \bibinfo{person}{Leen
  Lambers}.} \bibinfo{year}{2016}\natexlab{}.
\newblock \showarticletitle{On the Operationalization of Graph Queries with
  Generalized Discrimination Networks}. In \bibinfo{booktitle}{\emph{Graph
  Transformation - 9th International Conference, {ICGT} 2016, Vienna, Austria,
  July 5-6, 2016, Proceedings}} \emph{(\bibinfo{series}{Lecture Notes in
  Computer Science}, Vol.~\bibinfo{volume}{9761})},
  \bibfield{editor}{\bibinfo{person}{Rachid Echahed} {and}
  \bibinfo{person}{Mark Minas}} (Eds.). \bibinfo{publisher}{Springer},
  \bibinfo{pages}{170--186}.
\newblock
\urldef\tempurl%
\url{https://doi.org/10.1007/978-3-319-40530-8\_11}
\showDOI{\tempurl}


\bibitem[\protect\citeauthoryear{Biermann, Ehrig, Ermel, Golas, and
  Taentzer}{Biermann et~al\mbox{.}}{2010}]%
        {DBLP:conf/birthday/BiermannEEGT10}
\bibfield{author}{\bibinfo{person}{Enrico Biermann}, \bibinfo{person}{Hartmut
  Ehrig}, \bibinfo{person}{Claudia Ermel}, \bibinfo{person}{Ulrike Golas},
  {and} \bibinfo{person}{Gabriele Taentzer}.} \bibinfo{year}{2010}\natexlab{}.
\newblock \showarticletitle{Parallel Independence of Amalgamated Graph
  Transformations Applied to Model Transformation}. In
  \bibinfo{booktitle}{\emph{Graph Transformations and Model-Driven Engineering
  - Essays Dedicated to Manfred Nagl on the Occasion of his 65th Birthday}}
  \emph{(\bibinfo{series}{Lecture Notes in Computer Science},
  Vol.~\bibinfo{volume}{5765})}, \bibfield{editor}{\bibinfo{person}{Gregor
  Engels}, \bibinfo{person}{Claus Lewerentz}, \bibinfo{person}{Wilhelm
  Sch{\"{a}}fer}, \bibinfo{person}{Andy Sch{\"{u}}rr}, {and}
  \bibinfo{person}{Bernhard Westfechtel}} (Eds.).
  \bibinfo{publisher}{Springer}, \bibinfo{pages}{121--140}.
\newblock
\urldef\tempurl%
\url{https://doi.org/10.1007/978-3-642-17322-6\_7}
\showDOI{\tempurl}


\bibitem[\protect\citeauthoryear{Blair, Bencomo, and France}{Blair
  et~al\mbox{.}}{2009}]%
        {Blair+2009}
\bibfield{author}{\bibinfo{person}{Gordon Blair}, \bibinfo{person}{Nelly
  Bencomo}, {and} \bibinfo{person}{Robert~B. France}.}
  \bibinfo{year}{2009}\natexlab{}.
\newblock \showarticletitle{Models@run.time}.
\newblock \bibinfo{journal}{\emph{Computer}} \bibinfo{volume}{42},
  \bibinfo{number}{10} (\bibinfo{year}{2009}), \bibinfo{pages}{22--27}.
\newblock
\urldef\tempurl%
\url{https://doi.org/10.1109/MC.2009.326}
\showDOI{\tempurl}


\bibitem[\protect\citeauthoryear{B{\'{u}}r, Szil{\'{a}}gyi, V{\"{o}}r{\"{o}}s,
  and Varr{\'{o}}}{B{\'{u}}r et~al\mbox{.}}{2020}]%
        {DBLP:journals/sttt/BurSVV20}
\bibfield{author}{\bibinfo{person}{M{\'{a}}rton B{\'{u}}r},
  \bibinfo{person}{G{\'{a}}bor Szil{\'{a}}gyi}, \bibinfo{person}{Andr{\'{a}}s
  V{\"{o}}r{\"{o}}s}, {and} \bibinfo{person}{D{\'{a}}niel Varr{\'{o}}}.}
  \bibinfo{year}{2020}\natexlab{}.
\newblock \showarticletitle{Distributed graph queries over models@run.time for
  runtime monitoring of cyber-physical systems}.
\newblock \bibinfo{journal}{\emph{Int. J. Softw. Tools Technol. Transf.}}
  \bibinfo{volume}{22}, \bibinfo{number}{1} (\bibinfo{year}{2020}),
  \bibinfo{pages}{79--102}.
\newblock
\urldef\tempurl%
\url{https://doi.org/10.1007/s10009-019-00531-5}
\showDOI{\tempurl}


\bibitem[\protect\citeauthoryear{Búr, Szilágyi, Vörös, and Varró}{Búr
  et~al\mbox{.}}{2018}]%
  {Bur_2018_DistributedGraphQueriesforRuntimeMonitoringofCyberPhysicalSystems}
\bibfield{author}{\bibinfo{person}{Márton Búr}, \bibinfo{person}{Gábor
  Szilágyi}, \bibinfo{person}{András Vörös}, {and} \bibinfo{person}{Dániel
  Varró}.} \bibinfo{year}{2018}\natexlab{}.
\newblock \showarticletitle{Distributed {Graph} {Queries} for {Runtime}
  {Monitoring} of {Cyber}-{Physical} {Systems}}. In
  \bibinfo{booktitle}{\emph{Fundamental {Approaches} to {Software}
  {Engineering}}} \emph{(\bibinfo{series}{Lecture {Notes} in {Computer}
  {Science}})}. \bibinfo{publisher}{Springer, Cham}, \bibinfo{pages}{111--128}.
\newblock
\showISBNx{978-3-319-89362-4 978-3-319-89363-1}
\urldef\tempurl%
\url{https://doi.org/10.1007/978-3-319-89363-1_7}
\showDOI{\tempurl}


\bibitem[\protect\citeauthoryear{Catarinucci, Donno, Mainetti, Palano, Patrono,
  Stefanizzi, and Tarricone}{Catarinucci et~al\mbox{.}}{2015}]%
        {DBLP:journals/iotj/CatarinucciDMPP15}
\bibfield{author}{\bibinfo{person}{Luca Catarinucci},
  \bibinfo{person}{Danilo~De Donno}, \bibinfo{person}{Luca Mainetti},
  \bibinfo{person}{Luca Palano}, \bibinfo{person}{Luigi Patrono},
  \bibinfo{person}{Maria~Laura Stefanizzi}, {and} \bibinfo{person}{Luciano
  Tarricone}.} \bibinfo{year}{2015}\natexlab{}.
\newblock \showarticletitle{An IoT-Aware Architecture for Smart Healthcare
  Systems}.
\newblock \bibinfo{journal}{\emph{{IEEE} Internet of Things Journal}}
  \bibinfo{volume}{2}, \bibinfo{number}{6} (\bibinfo{year}{2015}),
  \bibinfo{pages}{515--526}.
\newblock
\urldef\tempurl%
\url{https://doi.org/10.1109/JIOT.2015.2417684}
\showDOI{\tempurl}


\bibitem[\protect\citeauthoryear{Combi, Gambini, Migliorini, and
  Posenato}{Combi et~al\mbox{.}}{2012}]%
        {Combi_2012_ModellingTemporalDatacentricMedicalProcesses}
\bibfield{author}{\bibinfo{person}{Carlo Combi}, \bibinfo{person}{Mauro
  Gambini}, \bibinfo{person}{Sara Migliorini}, {and} \bibinfo{person}{Roberto
  Posenato}.} \bibinfo{year}{2012}\natexlab{}.
\newblock \showarticletitle{Modelling {Temporal}, {Data}-centric {Medical}
  {Processes}}. In \bibinfo{booktitle}{\emph{Proceedings of the {2Nd} {ACM}
  {SIGHIT} {International} {Health} {Informatics} {Symposium}}}
  \emph{(\bibinfo{series}{{IHI} '12})}. \bibinfo{publisher}{ACM},
  \bibinfo{address}{New York, NY, USA}, \bibinfo{pages}{141--150}.
\newblock
\showISBNx{978-1-4503-0781-9}
\urldef\tempurl%
\url{https://doi.org/10.1145/2110363.2110382}
\showDOI{\tempurl}
\newblock
\shownote{event-place: Miami, Florida, USA.}


\bibitem[\protect\citeauthoryear{Cuadrado and de~Lara}{Cuadrado and
  de~Lara}{2013}]%
        {DBLP:conf/icmt/CuadradoL13}
\bibfield{author}{\bibinfo{person}{Jes{\'{u}}s~S{\'{a}}nchez Cuadrado} {and}
  \bibinfo{person}{Juan de Lara}.} \bibinfo{year}{2013}\natexlab{}.
\newblock \showarticletitle{Streaming Model Transformations: Scenarios,
  Challenges and Initial Solutions}. In \bibinfo{booktitle}{\emph{Theory and
  Practice of Model Transformations - 6th International Conference, ICMT@STAF
  2013, Budapest, Hungary, June 18-19, 2013. Proceedings}}
  \emph{(\bibinfo{series}{Lecture Notes in Computer Science},
  Vol.~\bibinfo{volume}{7909})}, \bibfield{editor}{\bibinfo{person}{Keith
  Duddy} {and} \bibinfo{person}{Gerti Kappel}} (Eds.).
  \bibinfo{publisher}{Springer}, \bibinfo{pages}{1--16}.
\newblock
\urldef\tempurl%
\url{https://doi.org/10.1007/978-3-642-38883-5\_1}
\showDOI{\tempurl}


\bibitem[\protect\citeauthoryear{DiCiccio and Efron}{DiCiccio and
  Efron}{1996}]%
        {diciccio1996bootstrap}
\bibfield{author}{\bibinfo{person}{Thomas~J DiCiccio} {and}
  \bibinfo{person}{Bradley Efron}.} \bibinfo{year}{1996}\natexlab{}.
\newblock \showarticletitle{Bootstrap confidence intervals}.
\newblock \bibinfo{journal}{\emph{Statistical science}} (\bibinfo{year}{1996}),
  \bibinfo{pages}{189--212}.
\newblock


\bibitem[\protect\citeauthoryear{Dou, Bianculli, and Briand}{Dou
  et~al\mbox{.}}{2017}]%
  {Dou_2017_AModelDrivenApproachtoTraceCheckingofPatternBasedTemporalProperties}
\bibfield{author}{\bibinfo{person}{Wei Dou}, \bibinfo{person}{Domenico
  Bianculli}, {and} \bibinfo{person}{Lionel Briand}.}
  \bibinfo{year}{2017}\natexlab{}.
\newblock \showarticletitle{A {Model}-{Driven} {Approach} to {Trace} {Checking}
  of {Pattern}-{Based} {Temporal} {Properties}}. In
  \bibinfo{booktitle}{\emph{2017 {ACM}/{IEEE} 20th {International} {Conference}
  on {Model} {Driven} {Engineering} {Languages} and {Systems} ({MODELS})}}.
  \bibinfo{publisher}{IEEE}, \bibinfo{address}{Austin, TX},
  \bibinfo{pages}{323--333}.
\newblock
\showISBNx{978-1-5386-3492-9}
\urldef\tempurl%
\url{https://doi.org/10.1109/MODELS.2017.9}
\showDOI{\tempurl}


\bibitem[\protect\citeauthoryear{Dávid, Ráth, and Varró}{Dávid
  et~al\mbox{.}}{2014}]%
        {David_2014_StreamingModelTransformationsByComplexEventProcessing}
\bibfield{author}{\bibinfo{person}{István Dávid}, \bibinfo{person}{István
  Ráth}, {and} \bibinfo{person}{Dániel Varró}.}
  \bibinfo{year}{2014}\natexlab{}.
\newblock \showarticletitle{Streaming {Model} {Transformations} {By} {Complex}
  {Event} {Processing}}. In \bibinfo{booktitle}{\emph{Model-{Driven}
  {Engineering} {Languages} and {Systems}}} \emph{(\bibinfo{series}{Lecture
  {Notes} in {Computer} {Science}})}. \bibinfo{publisher}{Springer, Cham},
  \bibinfo{pages}{68--83}.
\newblock
\showISBNx{978-3-319-11652-5 978-3-319-11653-2}
\urldef\tempurl%
\url{https://doi.org/10.1007/978-3-319-11653-2_5}
\showDOI{\tempurl}
\newblock
\shownote{Citation Key Alias: 10.1007/978-3-319-11653-2\_5.}


\bibitem[\protect\citeauthoryear{Dávid, Ráth, and Varró}{Dávid
  et~al\mbox{.}}{2018}]%
  {David_2018_FoundationsforStreamingModelTransformationsbyComplexEventProcessing}
\bibfield{author}{\bibinfo{person}{István Dávid}, \bibinfo{person}{István
  Ráth}, {and} \bibinfo{person}{Dániel Varró}.}
  \bibinfo{year}{2018}\natexlab{}.
\newblock \showarticletitle{Foundations for {Streaming} {Model}
  {Transformations} by {Complex} {Event} {Processing}}.
\newblock \bibinfo{journal}{\emph{Software \& Systems Modeling}}
  \bibinfo{volume}{17}, \bibinfo{number}{1} (\bibinfo{date}{Feb.}
  \bibinfo{year}{2018}), \bibinfo{pages}{135--162}.
\newblock
\showISSN{1619-1366, 1619-1374}
\urldef\tempurl%
\url{https://doi.org/10.1007/s10270-016-0533-1}
\showDOI{\tempurl}


\bibitem[\protect\citeauthoryear{Ehrig, Prange, and Taentzer}{Ehrig
  et~al\mbox{.}}{2004}]%
        {DBLP:conf/gg/EhrigPT04}
\bibfield{author}{\bibinfo{person}{Hartmut Ehrig}, \bibinfo{person}{Ulrike
  Prange}, {and} \bibinfo{person}{Gabriele Taentzer}.}
  \bibinfo{year}{2004}\natexlab{}.
\newblock \showarticletitle{Fundamental Theory for Typed Attributed Graph
  Transformation}. In \bibinfo{booktitle}{\emph{Graph Transformations, Second
  International Conference, {ICGT} 2004, Rome, Italy, September 28 - October 2,
  2004, Proceedings}} \emph{(\bibinfo{series}{Lecture Notes in Computer
  Science}, Vol.~\bibinfo{volume}{3256})},
  \bibfield{editor}{\bibinfo{person}{Hartmut Ehrig}, \bibinfo{person}{Gregor
  Engels}, \bibinfo{person}{Francesco Parisi{-}Presicce}, {and}
  \bibinfo{person}{Grzegorz Rozenberg}} (Eds.). \bibinfo{publisher}{Springer},
  \bibinfo{pages}{161--177}.
\newblock
\showISBNx{3-540-23207-9}
\urldef\tempurl%
\url{https://doi.org/10.1007/978-3-540-30203-2\_13}
\showDOI{\tempurl}


\bibitem[\protect\citeauthoryear{Esfahani, Yuan, Canavera, and Malek}{Esfahani
  et~al\mbox{.}}{2016}]%
        {esfahani2016inferring}
\bibfield{author}{\bibinfo{person}{Naeem Esfahani}, \bibinfo{person}{Eric
  Yuan}, \bibinfo{person}{Kyle~R Canavera}, {and} \bibinfo{person}{Sam Malek}.}
  \bibinfo{year}{2016}\natexlab{}.
\newblock \showarticletitle{{Inferring software component interaction
  dependencies for adaptation support}}.
\newblock \bibinfo{journal}{\emph{ACM Transactions on Autonomous and Adaptive
  Systems (TAAS)}} \bibinfo{volume}{10}, \bibinfo{number}{4}
  (\bibinfo{year}{2016}), \bibinfo{pages}{1--32}.
\newblock


\bibitem[\protect\citeauthoryear{Foundation}{Foundation}{2020}]%
        {emf-website}
\bibfield{author}{\bibinfo{person}{Eclipse Foundation}.}
  \bibinfo{year}{2020}\natexlab{}.
\newblock \bibinfo{booktitle}{\emph{Eclipse Modeling Framework (EMF)}}.
\newblock
\urldef\tempurl%
\url{https://www.eclipse.org/modeling/emf/}
\showURL{%
Retrieved Aug 10, 2020 from \tempurl}


\bibitem[\protect\citeauthoryear{Garcia, Bencomo, Parra, and
  Garcia-Paucar}{Garcia et~al\mbox{.}}{2019}]%
        {Bencomo2019}
\bibfield{author}{\bibinfo{person}{Antonio Garcia}, \bibinfo{person}{Nelly
  Bencomo}, \bibinfo{person}{Juan Parra}, {and} \bibinfo{person}{Luis~H.
  Garcia-Paucar}.} \bibinfo{year}{2019}\natexlab{}.
\newblock \showarticletitle{Querying and annotating model histories with
  time-aware patterns}. In \bibinfo{booktitle}{\emph{{IEEE} / {ACM} 22nd
  international conference on model driven engineering languages and systems
  ({MODELS})}}, \bibfield{editor}{\bibinfo{person}{Marouane Kessentini},
  \bibinfo{person}{Tao Yue}, \bibinfo{person}{Alexander Pretschner},
  \bibinfo{person}{Sebastian Voss}, {and} \bibinfo{person}{Loli Burgue~no}}
  (Eds.).
\newblock
\urldef\tempurl%
\url{https://research.aston.ac.uk/en/publications/querying-and-annotating-model-histories-with-time-aware-patterns}
\showURL{%
\tempurl}


\bibitem[\protect\citeauthoryear{Garlan, Schmerl, and Cheng}{Garlan
  et~al\mbox{.}}{2009}]%
        {Garlan+2009}
\bibfield{author}{\bibinfo{person}{David Garlan}, \bibinfo{person}{Bradley
  Schmerl}, {and} \bibinfo{person}{Shang-Wen Cheng}.}
  \bibinfo{year}{2009}\natexlab{}.
\newblock \showarticletitle{{Software Architecture-Based Self-Adaptation}}.
\newblock In \bibinfo{booktitle}{\emph{Autonomic Computing and Networking}}.
  \bibinfo{publisher}{Springer}, \bibinfo{pages}{31--55}.
\newblock
\urldef\tempurl%
\url{http://dx.doi.org/10.1007/978-0-387-89828-5_2}
\showURL{%
\tempurl}


\bibitem[\protect\citeauthoryear{Ghahremani, Giese, and Vogel}{Ghahremani
  et~al\mbox{.}}{2017}]%
        {Ghahremani.2017.Efficient}
\bibfield{author}{\bibinfo{person}{Sona Ghahremani}, \bibinfo{person}{Holger
  Giese}, {and} \bibinfo{person}{Thomas Vogel}.}
  \bibinfo{year}{2017}\natexlab{}.
\newblock \showarticletitle{Efficient Utility-Driven Self-Healing Employing
  Adaptation Rules for Large Dynamic Architectures}. In
  \bibinfo{booktitle}{\emph{Proceedings of the 14th International Conference on
  Autonomic Computing (ICAC)}}.
\newblock


\bibitem[\protect\citeauthoryear{Ghahremani, Giese, and Vogel}{Ghahremani
  et~al\mbox{.}}{2020}]%
        {SG-TAAS}
\bibfield{author}{\bibinfo{person}{Sona Ghahremani}, \bibinfo{person}{Holger
  Giese}, {and} \bibinfo{person}{Thomas Vogel}.}
  \bibinfo{year}{2020}\natexlab{}.
\newblock \showarticletitle{{Improving Scalability and Reward of Utility-Driven
  Self-Healing for Large Dynamic Architectures}}.
\newblock \bibinfo{journal}{\emph{ACM Trans. Auton. Adapt. Syst.}}
  \bibinfo{volume}{14}, \bibinfo{number}{3} (\bibinfo{date}{February}
  \bibinfo{year}{2020}).
\newblock
\urldef\tempurl%
\url{https://doi.org/10.1145/3380965}
\showURL{%
\tempurl}


\bibitem[\protect\citeauthoryear{Giese, Hildebrandt, and Seibel}{Giese
  et~al\mbox{.}}{2009}]%
        {DBLP:journals/eceasst/GieseHS09}
\bibfield{author}{\bibinfo{person}{Holger Giese}, \bibinfo{person}{Stephan
  Hildebrandt}, {and} \bibinfo{person}{Andreas Seibel}.}
  \bibinfo{year}{2009}\natexlab{}.
\newblock \showarticletitle{Improved Flexibility and Scalability by
  Interpreting Story Diagrams}.
\newblock \bibinfo{journal}{\emph{{ECEASST}}}  \bibinfo{volume}{18}
  (\bibinfo{year}{2009}).
\newblock
\urldef\tempurl%
\url{https://doi.org/10.14279/tuj.eceasst.18.268}
\showDOI{\tempurl}


\bibitem[\protect\citeauthoryear{Giese, Maximova, Sakizloglou, and
  Schneider}{Giese et~al\mbox{.}}{2019}]%
        {Giese_2019_MetricTemporalGraphLogicoverTypedAttributedGraphs}
\bibfield{author}{\bibinfo{person}{Holger Giese}, \bibinfo{person}{Maria
  Maximova}, \bibinfo{person}{Lucas Sakizloglou}, {and} \bibinfo{person}{Sven
  Schneider}.} \bibinfo{year}{2019}\natexlab{}.
\newblock \showarticletitle{Metric {{Temporal Graph Logic}} over {{Typed
  Attributed Graphs}}}. In \bibinfo{booktitle}{\emph{Fundamental {{Approaches}}
  to {{Software Engineering}}}} \emph{(\bibinfo{series}{Lecture {{Notes}} in
  {{Computer Science}}})}, \bibfield{editor}{\bibinfo{person}{Reiner
  H{\"a}hnle} {and} \bibinfo{person}{Wil {van der Aalst}}} (Eds.).
  \bibinfo{publisher}{{Springer International Publishing}},
  \bibinfo{pages}{282--298}.
\newblock
\showISBNx{978-3-030-16722-6}


\bibitem[\protect\citeauthoryear{G{\'{o}}mez, Cabot, and Wimmer}{G{\'{o}}mez
  et~al\mbox{.}}{2018}]%
        {DBLP:conf/er/GomezCW18}
\bibfield{author}{\bibinfo{person}{Abel G{\'{o}}mez}, \bibinfo{person}{Jordi
  Cabot}, {and} \bibinfo{person}{Manuel Wimmer}.}
  \bibinfo{year}{2018}\natexlab{}.
\newblock \showarticletitle{TemporalEMF: {A} Temporal Metamodeling Framework}.
  In \bibinfo{booktitle}{\emph{Conceptual Modeling - 37th International
  Conference, {ER} 2018, Xi'an, China, October 22-25, 2018, Proceedings}}
  \emph{(\bibinfo{series}{Lecture Notes in Computer Science},
  Vol.~\bibinfo{volume}{11157})}, \bibfield{editor}{\bibinfo{person}{Juan
  Trujillo}, \bibinfo{person}{Karen~C. Davis}, \bibinfo{person}{Xiaoyong Du},
  \bibinfo{person}{Zhanhuai Li}, \bibinfo{person}{Tok~Wang Ling},
  \bibinfo{person}{Guoliang Li}, {and} \bibinfo{person}{Mong{-}Li Lee}} (Eds.).
  \bibinfo{publisher}{Springer}, \bibinfo{pages}{365--381}.
\newblock
\urldef\tempurl%
\url{https://doi.org/10.1007/978-3-030-00847-5\_26}
\showDOI{\tempurl}


\bibitem[\protect\citeauthoryear{Guava}{Guava}{2020}]%
        {Guava}
\bibfield{author}{\bibinfo{person}{Guava}.} \bibinfo{year}{2020}\natexlab{}.
\newblock \bibinfo{booktitle}{\emph{Google Core Libraries for Java}}.
\newblock
\urldef\tempurl%
\url{https://github.com/google/guava}
\showURL{%
Retrieved May 17, 2020 from \tempurl}


\bibitem[\protect\citeauthoryear{Habel and Pennemann}{Habel and
  Pennemann}{2009}]%
  {Habel_2009_Correctnessofhighleveltransformationsystemsrelativetonestedconditions}
\bibfield{author}{\bibinfo{person}{Annegret Habel} {and}
  \bibinfo{person}{Karl-Heinz Pennemann}.} \bibinfo{year}{2009}\natexlab{}.
\newblock \showarticletitle{Correctness of High-Level Transformation Systems
  Relative to Nested Conditions}.
\newblock \bibinfo{journal}{\emph{Math. Struct. Comput. Sci.}}
  \bibinfo{volume}{19}, \bibinfo{number}{2} (\bibinfo{year}{2009}),
  \bibinfo{pages}{245--296}.
\newblock


\bibitem[\protect\citeauthoryear{Hartmann, Fouquet, Jimenez, Rouvoy, and
  Traon}{Hartmann et~al\mbox{.}}{2017}]%
        {DBLP:conf/seke/0001FJRT17}
\bibfield{author}{\bibinfo{person}{Thomas Hartmann},
  \bibinfo{person}{Fran{\c{c}}ois Fouquet}, \bibinfo{person}{Matthieu Jimenez},
  \bibinfo{person}{Romain Rouvoy}, {and} \bibinfo{person}{Yves~Le Traon}.}
  \bibinfo{year}{2017}\natexlab{}.
\newblock \showarticletitle{Analyzing Complex Data in Motion at Scale with
  Temporal Graphs}. In \bibinfo{booktitle}{\emph{The 29th International
  Conference on Software Engineering and Knowledge Engineering, Wyndham
  Pittsburgh University Center, Pittsburgh, PA, USA, July 5-7, 2017}},
  \bibfield{editor}{\bibinfo{person}{Xudong He}} (Ed.).
  \bibinfo{publisher}{{KSI} Research Inc. and Knowledge Systems Institute
  Graduate School}, \bibinfo{pages}{596--601}.
\newblock
\urldef\tempurl%
\url{https://doi.org/10.18293/SEKE2017-048}
\showDOI{\tempurl}


\bibitem[\protect\citeauthoryear{Havelund and Peled}{Havelund and
  Peled}{2018}]%
  {Havelund_2018_EfficientRuntimeVerificationofFirstOrderTemporalProperties}
\bibfield{author}{\bibinfo{person}{Klaus Havelund} {and} \bibinfo{person}{Doron
  Peled}.} \bibinfo{year}{2018}\natexlab{}.
\newblock \showarticletitle{Efficient {{Runtime Verification}} of
  {{First}}-{{Order Temporal Properties}}}.
\newblock In \bibinfo{booktitle}{\emph{Model {{Checking Software}}}},
  \bibfield{editor}{\bibinfo{person}{Mar{\'i}a del~Mar Gallardo} {and}
  \bibinfo{person}{Pedro Merino}} (Eds.). Vol.~\bibinfo{volume}{10869}.
  \bibinfo{publisher}{{Springer International Publishing}},
  \bibinfo{address}{{Cham}}, \bibinfo{pages}{26--47}.
\newblock
\showISBNx{978-3-319-94110-3 978-3-319-94111-0}


\bibitem[\protect\citeauthoryear{Havelund, Peled, and Ulus}{Havelund
  et~al\mbox{.}}{2017}]%
        {DBLP:conf/fmcad/HavelundPU17}
\bibfield{author}{\bibinfo{person}{Klaus Havelund}, \bibinfo{person}{Doron
  Peled}, {and} \bibinfo{person}{Dogan Ulus}.} \bibinfo{year}{2017}\natexlab{}.
\newblock \showarticletitle{First order temporal logic monitoring with BDDs}.
  In \bibinfo{booktitle}{\emph{2017 Formal Methods in Computer Aided Design,
  {FMCAD} 2017, Vienna, Austria, October 2-6, 2017}},
  \bibfield{editor}{\bibinfo{person}{Daryl Stewart} {and}
  \bibinfo{person}{Georg Weissenbacher}} (Eds.). \bibinfo{publisher}{{IEEE}},
  \bibinfo{pages}{116--123}.
\newblock
\showISBNx{978-0-9835678-7-5}
\urldef\tempurl%
\url{https://doi.org/10.23919/FMCAD.2017.8102249}
\showDOI{\tempurl}


\bibitem[\protect\citeauthoryear{Hildebrandt}{Hildebrandt}{2014}]%
        {Hildebrandt14}
\bibfield{author}{\bibinfo{person}{Stephan Hildebrandt}.}
  \bibinfo{year}{2014}\natexlab{}.
\newblock \emph{\bibinfo{title}{On the performance and conformance of triple
  graph grammar implementations}}.
\newblock \bibinfo{thesistype}{Ph.D. Dissertation}. \bibinfo{school}{University
  of Potsdam}.
\newblock
\urldef\tempurl%
\url{http://d-nb.info/1054564477}
\showURL{%
\tempurl}


\bibitem[\protect\citeauthoryear{Kephart and Chess}{Kephart and Chess}{2003}]%
        {Kephart&Chess2003}
\bibfield{author}{\bibinfo{person}{Jeffrey~O. Kephart} {and}
  \bibinfo{person}{David Chess}.} \bibinfo{year}{2003}\natexlab{}.
\newblock \showarticletitle{The Vision of Autonomic Computing}.
\newblock \bibinfo{journal}{\emph{Computer}} \bibinfo{volume}{36},
  \bibinfo{number}{1} (\bibinfo{year}{2003}), \bibinfo{pages}{41--50}.
\newblock
\urldef\tempurl%
\url{http://portal.acm.org/citation.cfm?id=642200}
\showURL{%
\tempurl}


\bibitem[\protect\citeauthoryear{Koymans}{Koymans}{1990}]%
        {Koymans_1990_Specifyingrealtimepropertieswithmetrictemporallogic}
\bibfield{author}{\bibinfo{person}{Ron Koymans}.}
  \bibinfo{year}{1990}\natexlab{}.
\newblock \showarticletitle{Specifying Real-Time Properties with Metric
  Temporal Logic}.
\newblock \bibinfo{journal}{\emph{Real-Time Syst.}} \bibinfo{volume}{2},
  \bibinfo{number}{4} (\bibinfo{year}{1990}), \bibinfo{pages}{255--299}.
\newblock


\bibitem[\protect\citeauthoryear{Lanese, Bucchiarone, and Montesi}{Lanese
  et~al\mbox{.}}{2010}]%
        {Rule-based_SASLanese2010}
\bibfield{author}{\bibinfo{person}{Ivan Lanese}, \bibinfo{person}{Antonio
  Bucchiarone}, {and} \bibinfo{person}{Fabrizio Montesi}.}
  \bibinfo{year}{2010}\natexlab{}.
\newblock \showarticletitle{{A Framework for Rule-Based Dynamic Adaptation}}.
  In \bibinfo{booktitle}{\emph{Proceedings of the 5th International Conference
  on Trustworthly Global Computing}}. \bibinfo{publisher}{Springer-Verlag},
  \bibinfo{address}{Berlin, Heidelberg}, \bibinfo{pages}{284--300}.
\newblock


\bibitem[\protect\citeauthoryear{Magee and Kramer}{Magee and Kramer}{1996}]%
        {MageeKramer1996}
\bibfield{author}{\bibinfo{person}{Jeff Magee} {and} \bibinfo{person}{Jeff
  Kramer}.} \bibinfo{year}{1996}\natexlab{}.
\newblock \showarticletitle{Dynamic Structure in Software Architectures}. In
  \bibinfo{booktitle}{\emph{Proc. of the 4th Symposium on Foundations of
  Software Engineering}}. \bibinfo{publisher}{ACM}, \bibinfo{pages}{3--14}.
\newblock
\urldef\tempurl%
\url{https://doi.org/10.1145/239098.239104}
\showDOI{\tempurl}


\bibitem[\protect\citeauthoryear{Mannhardt and Blinde}{Mannhardt and
  Blinde}{2017}]%
        {DBLP:conf/emisa/MannhardtB17}
\bibfield{author}{\bibinfo{person}{Felix Mannhardt} {and} \bibinfo{person}{Daan
  Blinde}.} \bibinfo{year}{2017}\natexlab{}.
\newblock \showarticletitle{Analyzing the Trajectories of Patients with Sepsis
  using Process Mining}. In \bibinfo{booktitle}{\emph{Joint Proceedings of the
  Radar tracks at the 18th International Working Conference on Business Process
  Modeling, Development and Support (BPMDS), and the 22nd International Working
  Conference on Evaluation and Modeling Methods for Systems Analysis and
  Development (EMMSAD), and the 8th International Workshop on Enterprise
  Modeling and Information Systems Architectures {(EMISA)} co-located with the
  29th International Conference on Advanced Information Systems Engineering
  2017 (CAiSE 2017), Essen, Germany, June 12-13, 2017}}
  \emph{(\bibinfo{series}{{CEUR} Workshop Proceedings},
  Vol.~\bibinfo{volume}{1859})}, \bibfield{editor}{\bibinfo{person}{Jens
  Gulden}, \bibinfo{person}{Selmin Nurcan}, \bibinfo{person}{Iris
  Reinhartz{-}Berger}, \bibinfo{person}{Wided Gu{\'{e}}dria},
  \bibinfo{person}{Palash Bera}, \bibinfo{person}{S{\'{e}}rgio Guerreiro},
  \bibinfo{person}{Michael Fellmann}, {and} \bibinfo{person}{Matthias
  Weidlich}} (Eds.). \bibinfo{publisher}{CEUR-WS.org}, \bibinfo{pages}{72--80}.
\newblock
\urldef\tempurl%
\url{http://ceur-ws.org/Vol-1859/bpmds-08-paper.pdf}
\showURL{%
\tempurl}


\bibitem[\protect\citeauthoryear{Manogaran, Varatharajan, Lopez, Kumar,
  Sundarasekar, and Thota}{Manogaran et~al\mbox{.}}{2018}]%
        {MANOGARAN2018375}
\bibfield{author}{\bibinfo{person}{Gunasekaran Manogaran}, \bibinfo{person}{R.
  Varatharajan}, \bibinfo{person}{Daphne Lopez},
  \bibinfo{person}{Priyan~Malarvizhi Kumar}, \bibinfo{person}{Revathi
  Sundarasekar}, {and} \bibinfo{person}{Chandu Thota}.}
  \bibinfo{year}{2018}\natexlab{}.
\newblock \showarticletitle{{A new architecture of Internet of Things and big
  data ecosystem for secured smart healthcare monitoring and alerting system}}.
\newblock \bibinfo{journal}{\emph{Future Generation Computer Systems}}
  \bibinfo{volume}{82} (\bibinfo{year}{2018}), \bibinfo{pages}{375 -- 387}.
\newblock
\urldef\tempurl%
\url{http://www.sciencedirect.com/science/article/pii/S0167739X17305149}
\showURL{%
\tempurl}


\bibitem[\protect\citeauthoryear{Moreno, C'amara, Garlan, and Schmerl}{Moreno
  et~al\mbox{.}}{2015}]%
        {moreno_2015_ProactiveSelfAdaptation}
\bibfield{author}{\bibinfo{person}{Gabriel~A. Moreno}, \bibinfo{person}{Javier
  C'amara}, \bibinfo{person}{David Garlan}, {and} \bibinfo{person}{Bradley
  Schmerl}.} \bibinfo{year}{2015}\natexlab{}.
\newblock \showarticletitle{{Proactive Self-adaptation Under Uncertainty: A
  Probabilistic Model Checking Approach}}. In
  \bibinfo{booktitle}{\emph{Proceedings of the 2015 10th Joint Meeting on
  Foundations of Software Engineering}} \emph{(\bibinfo{series}{ESEC/FSE
  2015})}. \bibinfo{publisher}{ACM}, \bibinfo{address}{New York, NY, USA},
  \bibinfo{pages}{1--12}.
\newblock


\bibitem[\protect\citeauthoryear{Rensink}{Rensink}{2004}]%
        {Rensink_2004_Representingfirstorderlogicusinggraphs}
\bibfield{author}{\bibinfo{person}{Arend Rensink}.}
  \bibinfo{year}{2004}\natexlab{}.
\newblock \showarticletitle{Representing First-Order Logic Using Graphs}. In
  \bibinfo{booktitle}{\emph{{{ICGT}}}}, Vol.~\bibinfo{volume}{4}.
  \bibinfo{publisher}{{Springer}}, \bibinfo{pages}{319--335}.
\newblock


\bibitem[\protect\citeauthoryear{Rhodes, Evans, Alhazzani, Levy, Antonelli,
  Ferrer, Kumar, Sevransky, Sprung, Nunnally, et~al\mbox{.}}{Rhodes
  et~al\mbox{.}}{2017}]%
        {rhodes2017surviving}
\bibfield{author}{\bibinfo{person}{Andrew Rhodes}, \bibinfo{person}{Laura~E
  Evans}, \bibinfo{person}{Waleed Alhazzani}, \bibinfo{person}{Mitchell~M
  Levy}, \bibinfo{person}{Massimo Antonelli}, \bibinfo{person}{Ricard Ferrer},
  \bibinfo{person}{Anand Kumar}, \bibinfo{person}{Jonathan~E Sevransky},
  \bibinfo{person}{Charles~L Sprung}, \bibinfo{person}{Mark~E Nunnally},
  {et~al\mbox{.}}} \bibinfo{year}{2017}\natexlab{}.
\newblock \showarticletitle{Surviving sepsis campaign: international guidelines
  for management of sepsis and septic shock: 2016}.
\newblock \bibinfo{journal}{\emph{Intensive care medicine}}
  \bibinfo{volume}{43}, \bibinfo{number}{3} (\bibinfo{year}{2017}),
  \bibinfo{pages}{304--377}.
\newblock


\bibitem[\protect\citeauthoryear{Roy, Abidi, and Abidi}{Roy
  et~al\mbox{.}}{2017}]%
        {roy2017monitoring}
\bibfield{author}{\bibinfo{person}{Patrice~C Roy}, \bibinfo{person}{Samina~Raza
  Abidi}, {and} \bibinfo{person}{Syed Sibte~Raza Abidi}.}
  \bibinfo{year}{2017}\natexlab{}.
\newblock \showarticletitle{Monitoring Medication Adherence in Smart
  Environments in the Context of Patient Self-management A Knowledge-driven
  Approach}.
\newblock In \bibinfo{booktitle}{\emph{Smart Technologies in Healthcare}}.
  \bibinfo{publisher}{CRC Press}, \bibinfo{pages}{195--223}.
\newblock


\bibitem[\protect\citeauthoryear{Sakizloglou, Ghahremani, Brand, Barkowsky, and
  Giese}{Sakizloglou et~al\mbox{.}}{2020}]%
        {2020_towards_highly_scalable_runtime_models_with_history}
\bibfield{author}{\bibinfo{person}{Lucas Sakizloglou}, \bibinfo{person}{Sona
  Ghahremani}, \bibinfo{person}{Thomas Brand}, \bibinfo{person}{Matthias
  Barkowsky}, {and} \bibinfo{person}{Holger Giese}.}
  \bibinfo{year}{2020}\natexlab{}.
\newblock \showarticletitle{Towards Highly Scalable Runtime Models with
  History}. In \bibinfo{booktitle}{\emph{15th {IEEE/ACM} International
  Symposium on Software Engineering for Adaptive and Self-Managing Systems,
  SEAMS@ICSE 2020, Seoul, South Korea, October, 2020}}.
  \bibinfo{publisher}{{IEEE} Computer Society}.
\newblock
\urldef\tempurl%
\url{https://arxiv.org/abs/2004.03727}
\showURL{%
\tempurl}


\bibitem[\protect\citeauthoryear{Schneider, Maximova, Sakizloglou, and
  Giese}{Schneider et~al\mbox{.}}{2020a}]%
        {mtgl-journal}
\bibfield{author}{\bibinfo{person}{Sven Schneider}, \bibinfo{person}{Maria
  Maximova}, \bibinfo{person}{Lucas Sakizloglou}, {and} \bibinfo{person}{Holger
  Giese}.} \bibinfo{year}{2020}\natexlab{a}.
\newblock \showarticletitle{Formal Testing of Timed Graph Transformation
  Systems using Metric Temporal Graph Logic}.
\newblock \bibinfo{journal}{\emph{Int. J. Softw. Tools Technol. Transf.}}
  (\bibinfo{year}{2020}).
\newblock
\newblock
\shownote{To appear.}


\bibitem[\protect\citeauthoryear{Schneider, Sakizloglou, Maximova, and
  Giese}{Schneider et~al\mbox{.}}{2020b}]%
        {DBLP:conf/gg/SchneiderSMG20}
\bibfield{author}{\bibinfo{person}{Sven Schneider}, \bibinfo{person}{Lucas
  Sakizloglou}, \bibinfo{person}{Maria Maximova}, {and} \bibinfo{person}{Holger
  Giese}.} \bibinfo{year}{2020}\natexlab{b}.
\newblock \showarticletitle{Optimistic and Pessimistic On-the-fly Analysis for
  Metric Temporal Graph Logic}. In \bibinfo{booktitle}{\emph{Graph
  Transformation - 13th International Conference, {ICGT} 2020, Held as Part of
  {STAF} 2020, Bergen, Norway, June 25-26, 2020, Proceedings}}
  \emph{(\bibinfo{series}{Lecture Notes in Computer Science},
  Vol.~\bibinfo{volume}{12150})}, \bibfield{editor}{\bibinfo{person}{Fabio
  Gadducci} {and} \bibinfo{person}{Timo Kehrer}} (Eds.).
  \bibinfo{publisher}{Springer}, \bibinfo{pages}{276--294}.
\newblock
\urldef\tempurl%
\url{https://doi.org/10.1007/978-3-030-51372-6\_16}
\showDOI{\tempurl}


\bibitem[\protect\citeauthoryear{Sebesty{\'e}nov{\'a}}{Sebesty{\'e}nov{\'a}}{2007}]%
        {sebestyenova2007case}
\bibfield{author}{\bibinfo{person}{Jolana Sebesty{\'e}nov{\'a}}.}
  \bibinfo{year}{2007}\natexlab{}.
\newblock \showarticletitle{Case-based reasoning in agent-based decision
  support system}.
\newblock \bibinfo{journal}{\emph{Acta Polytechnica Hungarica}}
  \bibinfo{volume}{4}, \bibinfo{number}{1} (\bibinfo{year}{2007}),
  \bibinfo{pages}{127--138}.
\newblock


\bibitem[\protect\citeauthoryear{Service}{Service}{2020}]%
        {nhsRecords}
\bibfield{author}{\bibinfo{person}{United Kingdom National~Health Service}.}
  \bibinfo{year}{2020}\natexlab{}.
\newblock \bibinfo{booktitle}{\emph{Records Management Code of Practice for
  Health and Social Care 2016}}.
\newblock
\urldef\tempurl%
\url{https://digital.nhs.uk/data-and-information/looking-after-information/data-security-and-information-governance/codes-of-practice-for-handling-information-in-health-and-care/records-management-code-of-practice-for-health-and-social-care-2016}
\showURL{%
Retrieved May 17, 2020 from \tempurl}


\bibitem[\protect\citeauthoryear{Steinberg, Budinsky, Merks, and
  Paternostro}{Steinberg et~al\mbox{.}}{2008}]%
        {steinberg2008emf}
\bibfield{author}{\bibinfo{person}{Dave Steinberg}, \bibinfo{person}{Frank
  Budinsky}, \bibinfo{person}{Ed Merks}, {and} \bibinfo{person}{Marcelo
  Paternostro}.} \bibinfo{year}{2008}\natexlab{}.
\newblock \bibinfo{booktitle}{\emph{EMF: eclipse modeling framework}}.
\newblock \bibinfo{publisher}{Pearson Education}.
\newblock


\bibitem[\protect\citeauthoryear{Sz{\'{a}}rnyas, Izs{\'{o}}, R{\'{a}}th,
  Harmath, Bergmann, and Varr{\'{o}}}{Sz{\'{a}}rnyas et~al\mbox{.}}{2014}]%
        {DBLP:conf/models/SzarnyasIRHBV14}
\bibfield{author}{\bibinfo{person}{G{\'{a}}bor Sz{\'{a}}rnyas},
  \bibinfo{person}{Benedek Izs{\'{o}}}, \bibinfo{person}{Istv{\'{a}}n
  R{\'{a}}th}, \bibinfo{person}{D{\'{e}}nes Harmath},
  \bibinfo{person}{G{\'{a}}bor Bergmann}, {and} \bibinfo{person}{D{\'{a}}niel
  Varr{\'{o}}}.} \bibinfo{year}{2014}\natexlab{}.
\newblock \showarticletitle{IncQuery-D: {A} Distributed Incremental Model Query
  Framework in the Cloud}. In \bibinfo{booktitle}{\emph{Model-Driven
  Engineering Languages and Systems - 17th International Conference, {MODELS}
  2014, Valencia, Spain, September 28 - October 3, 2014. Proceedings}}
  \emph{(\bibinfo{series}{Lecture Notes in Computer Science},
  Vol.~\bibinfo{volume}{8767})},
  \bibfield{editor}{\bibinfo{person}{J{\"{u}}rgen Dingel},
  \bibinfo{person}{Wolfram Schulte}, \bibinfo{person}{Isidro Ramos},
  \bibinfo{person}{Silvia Abrah{\~{a}}o}, {and} \bibinfo{person}{Emilio
  Insfr{\'{a}}n}} (Eds.). \bibinfo{publisher}{Springer},
  \bibinfo{pages}{653--669}.
\newblock
\urldef\tempurl%
\url{https://doi.org/10.1007/978-3-319-11653-2\_40}
\showDOI{\tempurl}


\bibitem[\protect\citeauthoryear{Ujhelyi, Bergmann, Heged{\"{u}}s,
  Horv{\'{a}}th, Izs{\'{o}}, R{\'{a}}th, Szatm{\'{a}}ri, and
  Varr{\'{o}}}{Ujhelyi et~al\mbox{.}}{2015}]%
        {DBLP:journals/scp/UjhelyiBHHIRSV15}
\bibfield{author}{\bibinfo{person}{Zolt{\'{a}}n Ujhelyi},
  \bibinfo{person}{G{\'{a}}bor Bergmann}, \bibinfo{person}{{\'{A}}bel
  Heged{\"{u}}s}, \bibinfo{person}{{\'{A}}kos Horv{\'{a}}th},
  \bibinfo{person}{Benedek Izs{\'{o}}}, \bibinfo{person}{Istv{\'{a}}n
  R{\'{a}}th}, \bibinfo{person}{Zolt{\'{a}}n Szatm{\'{a}}ri}, {and}
  \bibinfo{person}{D{\'{a}}niel Varr{\'{o}}}.} \bibinfo{year}{2015}\natexlab{}.
\newblock \showarticletitle{EMF-IncQuery: An integrated development environment
  for live model queries}.
\newblock \bibinfo{journal}{\emph{Sci. Comput. Program.}}  \bibinfo{volume}{98}
  (\bibinfo{year}{2015}), \bibinfo{pages}{80--99}.
\newblock
\urldef\tempurl%
\url{https://doi.org/10.1016/j.scico.2014.01.004}
\showDOI{\tempurl}


\bibitem[\protect\citeauthoryear{Vogel and Giese}{Vogel and Giese}{2010}]%
        {VG10}
\bibfield{author}{\bibinfo{person}{Thomas Vogel} {and} \bibinfo{person}{Holger
  Giese}.} \bibinfo{year}{2010}\natexlab{}.
\newblock \showarticletitle{{Adaptation and Abstract Runtime Models}}. In
  \bibinfo{booktitle}{\emph{SEAMS'10}}. \bibinfo{publisher}{ACM},
  \bibinfo{pages}{39--48}.
\newblock
\urldef\tempurl%
\url{http://dx.doi.org/10.1145/1808984.1808989}
\showURL{%
\tempurl}


\bibitem[\protect\citeauthoryear{Vogel, Seibel, and Giese}{Vogel
  et~al\mbox{.}}{2010}]%
        {DBLP:conf/models/VogelSG10}
\bibfield{author}{\bibinfo{person}{Thomas Vogel}, \bibinfo{person}{Andreas
  Seibel}, {and} \bibinfo{person}{Holger Giese}.}
  \bibinfo{year}{2010}\natexlab{}.
\newblock \showarticletitle{The Role of Models and Megamodels at Runtime}. In
  \bibinfo{booktitle}{\emph{Models in Software Engineering - Workshops and
  Symposia at {MODELS} 2010, Oslo, Norway, October 2-8, 2010, Reports and
  Revised Selected Papers}} \emph{(\bibinfo{series}{Lecture Notes in Computer
  Science}, Vol.~\bibinfo{volume}{6627})},
  \bibfield{editor}{\bibinfo{person}{J{\"{u}}rgen Dingel} {and}
  \bibinfo{person}{Arnor Solberg}} (Eds.). \bibinfo{publisher}{Springer},
  \bibinfo{pages}{224--238}.
\newblock
\urldef\tempurl%
\url{https://doi.org/10.1007/978-3-642-21210-9\_22}
\showDOI{\tempurl}


\bibitem[\protect\citeauthoryear{Weyns and Calinescu}{Weyns and
  Calinescu}{2015}]%
        {DBLP:conf/icse/WeynsC15}
\bibfield{author}{\bibinfo{person}{Danny Weyns} {and} \bibinfo{person}{Radu
  Calinescu}.} \bibinfo{year}{2015}\natexlab{}.
\newblock \showarticletitle{Tele Assistance: {A} Self-Adaptive Service-Based
  System Exemplar}. In \bibinfo{booktitle}{\emph{10th {IEEE/ACM} International
  Symposium on Software Engineering for Adaptive and Self-Managing Systems,
  {SEAMS} 2015, Florence, Italy, May 18-19, 2015}},
  \bibfield{editor}{\bibinfo{person}{Paola Inverardi} {and}
  \bibinfo{person}{Bradley~R. Schmerl}} (Eds.). \bibinfo{publisher}{{IEEE}
  Computer Society}, \bibinfo{pages}{88--92}.
\newblock
\urldef\tempurl%
\url{https://doi.org/10.1109/SEAMS.2015.27}
\showDOI{\tempurl}


\end{thebibliography}
\end{document}